\documentclass[a4paper,11pt]{article}

\pdfoutput=1 

\usepackage{jcappub_ver}
\usepackage{amsmath}
\usepackage{graphicx}

\usepackage[english]{babel}
\usepackage[utf8]{inputenc}
\usepackage{pdflscape}
\usepackage{enumerate}
\usepackage{amsbsy}
\usepackage{amsmath} 
\usepackage{graphics}
\usepackage{mathrsfs}
\usepackage{wrapfig}
\usepackage{mathtools}
\usepackage{accents}

\usepackage{amsfonts}
\usepackage{pstricks}
\usepackage{color}
\usepackage{setspace}

\newcommand{\bea}{\begin{eqnarray}} \newcommand{\eea}{\end{eqnarray}}
\newcommand{\el}{\nonumber \\}
\newcommand{\re}[1]{(\ref{#1})}

\newcommand{\pat}{\partial}

\renewcommand{\sec}[1]{section \ref{#1}}

\renewcommand{\a}{\alpha}
\renewcommand{\b}{\beta}
\renewcommand{\c}{\gamma}
\renewcommand{\d}{\delta}
\newcommand{\e}{\epsilon}
\newcommand{\f}{\zeta}

\newcommand{\ha}{\frac{1}{2}}

\newcommand{\rmd}{\mathrm{d}}

\newcommand{\ie}{i.e.\ }

\newcommand{\half}{\ha}

\DeclareRobustCommand{\dR}{\accentset{\Delta}{R}} 

\allowdisplaybreaks

\title{Scalar fields with derivative coupling to curvature in the Palatini and the metric formulation}

\author[a]{Hamed Bouzari Nezhad}
\author[a, b]{and Syksy R\"{a}s\"{a}nen}

\affiliation[a]{University of Helsinki, Helsinki Institute of Physics,\\ P.O. Box 64, FIN-00014 University of Helsinki, Finland}

\affiliation[b]{University of Helsinki, Department of Physics,\\ P.O. Box 64, FIN-00014 University of Helsinki, Finland}

\emailAdd{hamed.bouzarinezhad@gmail.com}
\emailAdd{syksy.rasanen@iki.fi}

\abstract{
We study models where a scalar field has derivative and non-derivative couplings to the Ricci tensor and the co-Ricci tensor with a view to inflation. We consider both the metric formulation and the Palatini formulation. In the Palatini case, the couplings to the Ricci tensor and the Ricci scalar give the same result regardless of whether the connection is unconstrained or the non-metricity or the torsion is assumed to vanish. When the co-Ricci tensor is included, the unconstrained case and the zero torsion case are physically different. We reduce all the actions to the Einstein frame with minimally coupled matter, and find the leading order differences between the metric case and the Palatini cases.
}

\begin{document}

\begin{flushleft}
	\hfill		 HIP-2023-11/TH \\
\end{flushleft}

\maketitle
  
\setcounter{tocdepth}{2}

\setcounter{secnumdepth}{3}

\section{Introduction} \label{sec:intro}

Inflation is the most successful scenario for the early universe \cite{Starobinsky:1979ty, Starobinsky:1980te, Kazanas:1980tx, Guth:1980zm, Sato:1980yn, Mukhanov:1981xt, Linde:1981mu, Albrecht:1982wi, Hawking:1981fz, Chibisov:1982nx, Hawking:1982cz, Guth:1982ec, Starobinsky:1982ee, Sasaki:1986hm, Mukhanov:1988jd}, and its predictions agree well with observations \cite{Planck:2018jri}. The simplest candidate for driving inflation is a scalar field. The field may be non-minimally coupled to curvature, as such couplings are generated by loop corrections \cite{Callan:1970ze}. Direct coupling to the Ricci scalar is the key feature of Higgs inflation \cite{Bezrukov:2007, Bezrukov:2013, Bezrukov:2015, Rubio:2018ogq}. Derivatives of the field can also couple to curvature \cite{Capozziello:1999uwa,  Capozziello:1999xt, Daniel:2007kk, Sushkov:2009hk, Germani:2011mx, Tsujikawa:2012mk, Yang:2015pga, Fu:2019ttf, Sato:2020ghj}. In the Higgs case, inflationary models with such couplings are called New Higgs Inflation \cite{Germani:2010gm, Germani:2010ux, Germani:2014hqa, DiVita:2015bha, Escriva:2016cwl, Fumagalli:2017cdo, Granda:2019wip, Granda:2019wyi, Fumagalli:2020ody}. When both derivative and non-derivative non-minimal couplings are present, the theories are sometimes called hybrid models \cite{Kobayashi:2010cm, Kamada:2010qe, Kobayashi:2011nu, Kamada:2012se, Kamada:2013bia, Kunimitsu:2015faa, Sato:2017qau}.
 
Generic actions with derivative couplings to the curvature, like generic actions with higher order curvature terms, lead to higher than second order equations of motion, which involve extra degrees of freedom that suffer from the Ostrogradsky instability \cite{Woodard:2006nt}. The most general scalar-tensor theories with second order equations of motion, called Horndeski theories, are explicitly known \cite{Horndeski:1974wa, Langlois:2018dxi, Kobayashi:2019hrl}. They are, however, not the most general stable scalar-tensor theories, because it is possible that the theory is degenerate and some degrees of freedom are not physical. On the gravity side, the simplest example is $f(R)$ theory \cite{Woodard:2006nt}. Degenerate higher order scalar-tensor theories (DHOST) have been explicitly catalogued up to terms cubic in the second derivatives of the field \cite{Langlois:2018dxi, Kobayashi:2019hrl}. The only such theories that are phenomenologically viable (with propagating gravitational waves and a Newtonian limit), at least at linear order in perturbation theory, are those that are related to Horndeski theories by an invertible disformal transformation \cite{Langlois:2017mxy} (see also \cite{Takahashi:2021ttd, Takahashi:2022mew}). Beyond DHOST are U-degenerate scalar-tensor theories, which are degenerate only in the unitary gauge, where the gradient of the scalar field has to be timelike \cite{Langlois:2015cwa, DeFelice:2018ewo, DeFelice:2021hps, Joshi:2021azw, Joshi:2023otx, Takahashi:2023jro}. They have also been explicitly catalogued up to third order in second derivatives, and the procedure for determining whether a theory with arbitrary powers of second derivatives is DHOST or U-degenerate or neither is known.

These results are for the metric formulation of gravity. In other formulations that are equivalent for the Einstein--Hilbert action with minimally coupled matter but physically distinct for more complicated actions, the stability properties of non-minimally coupled scalar fields have not been completely categorised. (For Horndeski theories in teleparallel and symmetric teleparallel gravity, see \cite{Bahamonde:2019shr, Bahamonde:2022cmz}.) One of the most common alternatives to the metric formulation is the Palatini formulation, where the connection is an independent variable \cite{einstein1925, ferraris1982}.\footnote{Some works have taken the metric formulation Horndeski action and simply replaced the Levi--Civita connection with a connection treated as an independent variable. In general, the resulting theories do not have second order equations of motion and are not stable \cite{Helpin:2019kcq, Helpin:2019vrv, Dong:2021jtd, Dong:2022cvf}.} Higgs inflation, where the field couples directly to the Ricci scalar has been much studied in the Palatini formulation, and the predictions are different than in the metric case \cite{Bauer:2008, Bauer:2010, Rasanen:2017, Racioppi:2017spw, Markkanen:2017tun, Enckell:2018a, Rasanen:2018fom, Rasanen:2018ihz, Rubio:2019, Jinno:2019und, Tenkanen:2020dge, Shaposhnikov:2020fdv, McDonald:2020lpz, Shaposhnikov:2020frq, Enckell:2020lvn, Antoniadis:2021axu, Mikura:2021clt, Ito:2021ssc, Karananas:2022byw, Dux:2022kuk, Gialamas:2023flv, Piani:2023aof, Poisson:2023tja}. Inflation in the case when derivatives of the field couple directly to the curvature has also been studied \cite{Gumjudpai:2016ioy, Galtsov:2018xuc, Gialamas:2020vto, Dioguardi:2023jwa}; in \cite{Luo:2014eda}, such a theory was used for quintessence (see also \cite{Aoki:2018lwx, Aoki:2019rvi, Galtsov:2020jnu, BeltranJimenez:2020sqf}). Unlike in the case when only the field couples directly to the curvature, in the derivative coupling case the results of the metric and the Palatini formulation are close to each other. We extend previous work by including the co-Ricci tensor in the cases when the connection is taken to be metric-compatible or torsion-free a priori. When parts of the connection are constrained in the action, the theory is in general different from the unconstrained case. For example, the theory with an Einstein--Hilbert term plus a term quadratic term in the antisymmetric part of the Ricci tensor is stable in the zero torsion case, but unstable in the unconstrained case \cite{BeltranJimenez:2019acz, Annala:2022gtl}.

In \sec{sec:calc} we give the geometrical background for the Palatini formulation and present the action. We shift to the Einstein frame with minimally coupled matter by making a disformal transformation followed by solving the remaining pieces of the connection from the equation of motion and inserting them back into the action. We calculate the leading order differences between the Palatini cases when the connection is unconstrained, when non-metricity or torsion is put to zero, and the metric case. In \sec{sec:conc} we summarise our findings and outline open questions. Some technical details are relegated to appendices \ref{app:con} and \ref{app:Riemann}.

\section{Non-minimal coupling to kinetic terms} \label{sec:calc}

\subsection{Curvature, non-metricity, and torsion}

In the Palatini formulation the metric $g_{\a\b}$ and the connection $\Gamma^\c_{\a\b}$ are independent variables. The connection, defined with the covariant derivative as $\nabla_\b A^\a=\pat_\b A^\a + \Gamma^\a_{\b\c} A^\c$, can be decomposed as
\bea \label{Gamma}
  \Gamma^\c_{\a\b} &=& \mathring\Gamma^\c_{\a\b} + L^\c{}_{\a\b} = \mathring\Gamma^\c_{\a\b} + J^\c{}_{\a\b} + K^\c{}_{\a\b} \ ,
\eea
where $\mathring\Gamma_{\a\b}^\c$ is the Levi--Civita connection of the metric $g_{\a\b}$. We denote quantities defined with the Levi--Civita connection by $\mathring{}$. In the second equality we have decomposed the distortion tensor $L^\c{}_{\a\b}$ into the disformation tensor $J_{\a\b\c}$ and the contortion tensor $K_{\a\b\c}$, defined as
\bea \label{L}
  J_{\a\b\c} &\equiv& \frac{1}{2} \left(Q_{\a\b\c}  - Q_{\c\a\b} - Q_{\b\a\c} \right) \ , \qquad K_{\a\b\c} \equiv \ha (T_{\a\b\c} + T_{\c\a\b} + T_{\b\a\c} ) \ ,
\eea
where $Q_{\a\b\c}$ and $T_{\a\b\c}$ are the non-metricity and the torsion, respectively, defined as
 \bea \label{TQ}
  \qquad Q_{\c\a\b} \equiv \nabla_\c g_{\a\b}  \ , \qquad T^\c{}_{\a\b} &\equiv& 2 \Gamma^{\c}_{[\a\b]} \ .
\eea
We have $Q_{\c\a\b}=Q_{\c(\a\b)}$, $J_{\a\b\c}=J_{\a(\b\c)}$, and $K^\c{}_\a{}^\b=K^{[\c}{}_\a{}^{\b]}$.

The Riemann tensor can be decomposed into the Levi--Civita and the distortion contributions as
\bea \label{Riemann}
  R^{\a}{}_{\b\c\d} = \mathring R^{\a}{}_{\b\c\d} + 2 \mathring \nabla_{[\c} L^\a{}_{\d]\b} + 2 L^\a{}_{[\c|\mu|} L^\mu{}_{\d]\b} \ .
\eea
There are three independent first contractions of the Riemann tensor, called Ricci-type tensors,
\begin{eqnarray}\label{RicciTensors}
  R_{\a\b}\equiv R^{\c}{}_{\a\c\b} \ , \quad
  \hat{R}_{\a\b}\equiv g_{\a\e}g^{\c\d}R^{\e}{}_{\c\d\b} \ , \quad
  \tilde{R}_{\a\b}\equiv R^{\c}{}_{\c\a\b} \ .
\end{eqnarray}
The first is the Ricci tensor, the second is the co-Ricci tensor, and the third is the homothetic curvature tensor. There is only one independent Ricci scalar, $R=-\hat R$, $\tilde{R}=0$. Instead of the co-Ricci tensor, it can be convenient to use the average of the co-Ricci tensor and the Ricci tensor. Using the definition \re{RicciTensors} and the decompositions \re{Gamma}, \re{L}, and \re{Riemann}, we see that the average vanishes when $Q_{\a\b\c}=0$,
\begin{equation} \label{ricciav}
	\dR_{\a\b} \equiv \half (\hat R_{\a\b}+R_{\a\b} ) = g^{\mu\nu}\nabla_{[\b} Q_{\mu]\nu\a} - \ha T^{\mu\nu}{}_\b Q_{\mu\nu\a} \ . 
\end{equation}
The Einstein tensor is
\bea \label{Einstein}
  G_{\a\b} &\equiv& - \frac{1}{4} \epsilon_{\a\c}{}^{\mu_1\nu_1} \epsilon_\b{}^{\c\mu_2\nu_2} R_{\mu_2\nu_2\mu_1\nu_1} = \ha (R_{\a\b} - \hat R_{\a\b} - g_{\a\b} R ) \ ,
\eea
where $\epsilon_{\a\b\c\d}$ is the Levi--Civita tensor.

\subsection{The action} \label{sec:action}

We consider a scalar field $\varphi$ whose kinetic term $X_{\a\b}\equiv\pat_\a \varphi \pat_\b \varphi$ couples linearly to the first traces of the Riemann tensor, while $\varphi$ can appear non-linearly. (General non-linear couplings have been studied in \cite{Annala:2022gtl}.) The homothetic curvature tensor does not appear because it is antisymmetric, so in the Palatini case, we have couplings to $R_{\a\b}$, $\hat R_{\a\b}$ and $R$, and the action is
\bea \label{action1}
  S &=& \int \rmd^4 x \sqrt{-g} \bigg[ \ha F(\varphi) g^{\a\b} R_{\a\b} - \ha K(\varphi) g^{\a\b} X_{\a\b} + \ha \a_1(\varphi) g^{\a\b} g^{\c\d} R_{\a\b} X_{\c\d} \el
  && + \ha \a_2(\varphi) g^{\a\c} g^{\b\d} R_{\a\b} X_{\c\d} + \ha \a_3(\varphi) g^{\b\c} g^{\d\mu} R^\a{}_{\b\c\d} X_{\a\mu} - V(\varphi) + \mathcal{L}_{\textrm{m}}(\Psi, \varphi, g^{\a\b}) \bigg] \el
  &=& \int \rmd^4 x \sqrt{-g} \left[ \ha ( F + \a_1 X ) R - \ha K X + \ha ( \a_2 R^{\a\b} + \a_3 \hat R^{\a\b} ) X_{\a\b} - V + \mathcal{L}_{\textrm{m}} \right] \ ,
\eea
where $g=\det g_{\a\b}$, $X\equiv g^{\a\b} X_{\a\b}$, and $\mathcal{L}_{\textrm{m}}(\Psi,\varphi,g_{\a\b})$ is a matter action\footnote{Fermion kinetic terms involve the connection. We neglect them; it is always possible to assume that they couple only to the Levi--Civita connection, and thus do not contribute to the distortion.}, with $\Psi$ denoting any matter degrees of freedom other than $\varphi$.

In the metric case $\hat R_{\a\b}=-R_{\a\b}$, so we can put $\a_3=0$. Then when $\a_1=-\ha\a_2$, the action is of the Horndeski form, and there are no extra degrees of freedom, otherwise there is an extra ghost \cite{Horndeski:1974wa}. If also $\a_2>0$, the scalar degree of freedom corresponding to $\varphi$ is healthy, otherwise it is a ghost \cite{Germani:2010gm}.

In the Palatini case, the theory is different depending on which, if any, constraints are imposed on the connection. The case without constraints has been studied in \cite{Aoki:2018lwx, Aoki:2019rvi}. Solving the connection equation obtained by varying \re{action1} with respect to $\Gamma^\c_{\a\b}$ and inserting the solution into the action gives a metric theory with a modified scalar sector. For an action including \re{action1} but more general, it was shown in \cite{Aoki:2019rvi} that the theory is at least U-degenerate (and can be DHOST or Horndeski). The reason is that it is symmetric under the projective transformation $\Gamma^\c_{\a\b} \to \Gamma^\c_{\a\b} + \delta^\c{}_\b V_\a$, where $V_\a$ is an arbitrary vector field. When the gradient of the scalar field is timelike, the ghost is subsumed in the unphysical projective mode.\footnote{In \cite{DeFelice:2018ewo} it is argued that U-degenerate theories could be healthy. However, it is not clear how the theory behaves when spatial gradients are larger than the time derivatives \cite{DeFelice:2021hps}, for example during reheating or close to the vacuum at late times. In general, projective symmetry does not guarantee the absence of ghosts, and whether ghosts appear can depend on the background \cite{Annala:2022gtl}.} The results of \cite{Aoki:2019rvi} show that for the action \re{action1}, the theory is in the DHOST class. (For the case when $X_{\a\b}$ couples only to the Ricci scalar and the Einstein tensor \re{Einstein}, \ie $\a_3=-\a_2$, this was shown already in \cite{Aoki:2018lwx}.)

We will consider the case when either non-metricity or torsion is set to zero.

\subsection{Disformal transformation} \label{sec:dis}

We could solve the connection separately in the cases with zero non-metricity or zero torsion and insert the solution back into the action. However, it is easier to first get rid of all non-minimal couplings except those to $\dR_{\a\b}$ with a disformal transformation. This will also establish that the result is the same in the case when the connection is unconstrained and when non-metricity is put to zero, and that in the zero torsion case the difference arises only from $\dR_{\a\b}$. It has been shown that observables such as inflationary power spectra are invariant under disformal transformations at least for Horndeski theories \cite{Minamitsuji:2014waa, Tsujikawa:2014uza, Watanabe:2015uqa, Motohashi:2015pra, Domenech:2015hka} (see also \cite{Chiba:2020mte}).

We will perform an invertible disformal transformation in the action \re{action1} such that only a coupling to $\dR_{\a\b}$ remains \cite{Bettoni:2013diz, Zumalacarregui:2013pma,Minamitsuji:2014waa,Tsujikawa:2014uza,  Watanabe:2015uqa, Motohashi:2015pra, Domenech:2015hka,Fumagalli:2016afy,Takahashi:2017zgr,Chiba:2020mte,Annala:2022gtl}:
\bea \label{disformal}
  g_{\a\b} &=& \gamma_1(\varphi, \tilde X) \tilde g_{\a\b} + \gamma_2(\varphi, \tilde X) X_{\a\b} \ ,
\eea
where $\tilde X\equiv \tilde g^{\a\b} X_{\a\b}$. The inverse transformation is
\bea \label{tildeg}
  \tilde g_{\a\b} &=& \tilde\gamma_1(\varphi, X) g_{\a\b} + \tilde\gamma_2(\varphi, X) X_{\a\b} \ .
\eea
The original and tilded transformation functions are related to each other as $\tilde\c_1=1/\c_1$, $\tilde\c_2=-\c_2/\c_1$. The inverse metric is
\begin{equation}\label{tildeginv}
  g^{\a\b} = \frac{1}{\gamma_1} \tilde g^{\a\b} - \frac{\gamma_2}{\gamma_1(\gamma_1+\gamma_2 \tilde X)} \tilde g^{\a\mu}\tilde g^{\b\nu} X_{\mu\nu} \ ,
\end{equation}
and $\tilde g^{\a\b}$ is given by the same expression with the replacements $\c_i\to\tilde\c_i$, $\tilde X\to X$, $\tilde g^{\a\b}\to g^{\a\b}$. These equations give us the relation between $X$ and $\tilde X$
\begin{equation} \label{X}
  X = \frac{\tilde X}{\gamma_1+ \gamma_2 \tilde X} \ .
\end{equation}
As the original and tilded variables are in a symmetric position, $X$ as a function of $\tilde X$ is, again, given by the same equation with the original and tilded quantities switched. The determinants of the metrics are related by
\begin{equation} \label{detg}
	g = \tilde g \gamma_1^3 (\gamma_1 + \gamma_2 \tilde X ) \ .
\end{equation}

Under the disformal transformation \re{disformal}, the curvature coupling terms in the action \re{action1} transform as follows
\bea \label{disR}
  \sqrt{-g} g^{\a\b} R_{\a\b} &=& \sqrt{-\tilde g} \c_1 ( 1 + \c \tilde X )^{1/2} \left( \tilde g^{\a\b} R_{\a\b} - \frac{\c}{1 + \c \tilde X} \tilde g^{\a\c} \tilde g^{\b\d} R_{\a\b} X_{\c\d} \right) \el
  \sqrt{-g} g^{\a\c} g^{\b\d} R_{\a\b} X_{\c\d} &=& \sqrt{-\tilde g} ( 1 + \c \tilde X )^{-3/2} \tilde g^{\a\c} \tilde g^{\b\d} R_{\a\b} X_{\c\d} \el
  \sqrt{-g} g^{\b\c} g^{\d\mu} R^\a{}_{\b\c\d} X_{\a\mu} &=& \sqrt{-\tilde g} ( 1 + \c \tilde X )^{-1/2} \tilde g^{\b\c} \tilde g^{\d\mu} R^\a{}_{\b\c\d} X_{\a\mu} \ ,
\eea
where $\c\equiv\c_2/\c_1$.

Applying the disformal transformation \re{disformal} to the action \re{action1}, using the above results, writing the co-Ricci tensor $\hat R_{\a\b}$ in terms of $\dR_{\a\b}$ defined in \re{ricciav}, and dropping the tildes on $g_{\a\b}$ and $X$, we get
\bea \label{ActionDTI}
  S & = & \int \rmd^4x \sqrt{-g} \Bigg\{ \ha ( 1 + \c X)^{1/2} \left( \c_1 F + \frac{\a_1 X}{1 + \c X} \right) R + \frac{\a_{3} }{(1 + \c X)^{1/2}} \dR^{\a\b} X_{\a\b} \el
  && + \ha \frac{1}{(1+ \c X)^{1/2}} \left[ - F \c_2 + \frac{ \a_2 - \a_3 - ( \a_1 + \a_3 ) \c X }{1 + \c X} \right] R^{\a\b} X_{\a\b} \el
  && - \frac{\c_1}{2 ( 1 + \c X)^{1/2}} K X - \gamma_{1}^2 ( 1 + \c X )^{1/2} V \el
  && + \c_1^{2} ( 1 + \c X)^{1/2} \mathcal{L}_{\textrm{m}} \left[ \Psi, \varphi, \frac{1}{\gamma_1} g^{\a\b} - \frac{\c}{\gamma_1 ( 1 + \c X )} g^{\a\mu} g^{\b\nu} X_{\mu\nu} \right] \Bigg\} \ .
\eea
The non-minimal couplings to $R$ and $R_{\a\b}$ are eliminated by choosing
\bea \label{DTFixingI}
  && ( 1 + \c X)^{1/2} \left( \c_1 F + \frac{\a_1 X}{1 + \c X} \right) = 1 \el
  && F \c_2 + \frac{ \a_2 - \a_3 - ( \a_1 + \a_3 ) \c X}{1 + \c X} = 0 \ .
\eea
From \re{DTFixingI} we can solve for $\gamma_{1}$ and $\gamma_{2}$ in closed form. The solutions are not very illuminating, so we do not write them down. For $\a_3=0$ they simplify; the case $\a_1=-\ha\a_2$, $\a_3=0$ is given in \cite{Gialamas:2020vto}.

The disformal transformation is invertible and the original and transformed metric describe the same physics when $\c_1>0$, $\c_2\geq0$, $\c_1+ \tilde X \c_2>0$, $\tilde\c_1- X \pat_{X} \tilde\c_1 - X^2 \pat_{X} \tilde\c_2\neq0$. These conditions set a limit on the values $X$ can take. This is a limitation of the disformal transformation. Large spatial gradients such as may occur during preheating may mean that the coefficient of the Ricci tensor is not positive, so that even in the case $\a_3=0$, the theory cannot be mapped to a minimally coupled Einstein frame with a disformal transformation. For study of slow-roll inflation in the super-Hubble regime, this is not a problem.

Inserting $\c_1$ and $\c_2$ back into the action \re{ActionDTI}, we get (dropping the matter Lagrangian)
\bea \label{ActionDTII}
S = \int \rmd^4x \sqrt{-g} \left( \frac{1}{2} R + \mathcal{G}_{1} \dR^{\a\b} X_{\a\b} - \frac{1}{2} \mathcal{G}_{2} K X - \mathcal{G}_{3} V \right) \ ,
\eea
where we have defined
\bea \label{GFunctions}
  \mathcal{G}_{1} \equiv \frac{\a_{3}}{(1 + \c X)^{1/2}} \ , \
  \mathcal{G}_{2} \equiv \frac{\gamma_{1}}{(1 + \c X)^{1/2}} \ , \
  \mathcal{G}_{3} \equiv \gamma_{1}^2 (1 + \c X)^{1/2} \ .
\eea

If $\a_3=0$, then $ \mathcal{G}_{1}=0$. In this case the Ricci scalar is the only term that contains the connection, so the connection equation of motion gives the Levi--Civita connection. (In the case when there are no a priori constraints on the connection, it is determined only up to a projective transformation.) Inserting it back into the action we obtain a metric theory with a minimally coupled scalar field. The physics related to the distortion has been shifted to the modifications of the scalar field kinetic term and potential (and the matter Lagrangian). In the Einstein frame all matter couples to the scalar field and its kinetic term. We have not assumed anything about the connection, showing that if the co-Ricci tensor does not appear in the action, the physics is the same whether we keep the connection unconstrained or put non-metricity or torsion to zero. This is also the case in Palatini $f(R)$ theory, which can be reduced to the Einstein--Hilbert plus minimally coupled matter form via field transformations \cite{Magnano:1987zz, Koga:1998un, Sotiriou:2008, Afonso:2017}. If the non-metricity is put to zero a priori, \re{ricciav} shows that $\hat R_{\a\b}=-R_{\a\b}$, so the co-Ricci tensor is not independent, and there is no $\a_3$ coupling.

In any case, if $\a_3=0$, the action \re{ActionDTII} reduces to
\bea \label{actiona3zero}
  S &=& \int \rmd^4 x \sqrt{-g} \left[ \ha g^{\a\b} R_{\a\b} - \ha \frac{\c_1}{(1+ \c X)^{1/2}} K X - \c_1^{2} ( 1 + \c X )^{1/2} V \right] \el
  &\simeq& \int \rmd^4 x \sqrt{-g} \Bigg\{ \ha \mathring R - \frac{K X}{2 F} \left[1 - \left(\a_{1} + \a_{2}\right) X\right] \el&& - \frac{V}{F^{2}} \left[1 - \left(2 \a_{1} + \frac{1}{2} \a_{2}\right) X + \left(\a_{1}^{2} + 2 \a_{1} \a_{2} + \frac{5}{8} \a_{2}^{2}\right) X^{2}\right] \Bigg\} \ ,
\eea
where in the second equality we have expanded to second order in $X_{\a\b}$. For $\a_1=-\ha\a_2$, $\a_3=0$ the result agrees with \cite{Gialamas:2020vto}.

The action \re{actiona3zero} is manifestly in the Horndeski class. We noted in \sec{sec:action} that based on the results of \cite{Aoki:2019rvi}, the action is of the DHOST form. However, as written in the introduction, the only viable DHOST theories (at least to cubic order in second derivatives) seem to be those that are related to Horndeski theories by an invertible disformal transformation. There is no physical difference between Horndeski and DHOST theories as regards physical degrees of freedom and stability.

\subsection{Zero torsion case with $\a_3\neq0$} \label{sec:zerot}

When $\a_3\neq0$ and the non-metricity is non-zero, we have to solve the connection equation of motion and insert the solution back into the action. Let us consider the case with zero torsion. (The case with no constraints was considered in \cite{Aoki:2019rvi}.) Varying the action \re{ActionDTII} with respect to the distortion tensor (taking into account that it is symmetric in the last two indices) gives the equation of motion
\bea \label{LEoM}
  0 & = & g_{\b\c} L^{\d}{}_{\d\a} - L_{(\b\c)\a} - L_{(\b|\a|\c)} + g_{\a(\b}L_{\c)}{}^{\d}{}_{\d} \el
  && + \mathcal{G}_{1} \big(L^{\d}{}_{\d\a} X_{\b\c} - g_{\b\c} L^{\d\e}{}_{\a} X_{\d\e} - X_{\a} Y_{\b\c} + g_{\b\c} \mathring{\nabla}_{\d}X_{\a}{}^{\d} + g_{\a(\b}\mathring{\nabla}^{\d}X_{\c)\d} + L_{(\b}{}^{\d}{}_{\c)}X_{\a\d} \el
  && - L_{(\b|\a|}{}^{\d}X_{\c)\d} - L_{(\b}{}^{\d}{}_{|\a|}X_{\c)\d} - L_{(\b}{}^{\d}{}_{|\d}X_{\a|\c)} + L^{\d}{}_{(\b|\a|}X_{\c)\d} - \frac{3}{2} \mathring{\nabla}_{\a}X_{\b\c} + g_{\a(\b}L_{\c)}{}^{\d\e}X_{\d\e}\big) \el
  && + \mathcal{G}_{1}' \big(X g_{\b\c} X_{\a} + X g_{\a(\b}X_{\c)} - 2 X_{\a} X_{\b\c}\big) \el
  && + \partial_{X}\mathcal{G}_{1} \big( g_{\b\c} X_{\a\d} \mathring{\nabla}^{\d}X + g_{\a(\b}X_{\c)}{}^{\d}\mathring{\nabla}_{\d}X - X_{\b\c} \mathring{\nabla}_{\a}X - X_{\a(\b}\mathring{\nabla}_{\c)}X \big) \ ,
\eea
where $X_\a\equiv\partial_\a\varphi$, $Y_{\a\b}\equiv\mathring\nabla_\a\mathring\nabla_\b\varphi$, and prime denotes partial derivative with respect to $\varphi$. The general solution has the form
\bea \label{LGuess}
  L_{\a\b\c} & = & l_{1} g_{\b\c} X_{\a} + l_{2} g_{\a(\b}X_{\c)} + l_{3} X_{\a} Y_{\b\c} + l_{4} \mathring{\nabla}_{\a}X_{\b\c} + l_{5} g_{\b\c} \mathring{\nabla}_{\a}X + l_{6} g_{\b\c} \mathring{\nabla}_{\d}X_{\a}{}^{\d} \el&& + l_{7} g_{\a(\b} \mathring{\nabla}^{\d}X_{\c)\d} + l_{8} g_{\a(\b} \mathring{\nabla}_{\c)}X + l_{9} X_{\a} X_{\b\c} + l_{10} X_{\b\c} \mathring{\nabla}_{\a}X + l_{11} X_{\b\c} \mathring{\nabla}_{\d}X_{\a}{}^{\d} \el&& + l_{12} g_{\b\c} X_{\a\d} \mathring{\nabla}^{\d}X + l_{13} X_{\a(\b} \mathring{\nabla}_{\c)}X + l_{14} g_{\a(\b}X_{\c)}{}^{\d} \mathring{\nabla}_{\d}X + l_{15} X_{\a\d} X_{\b\c} \mathring{\nabla}^{\d}X \ ,
\eea
Inserting \re{LGuess} into \re{LEoM}, we solve for the coefficients $l_i(\varphi, X)$. The result is rather lengthy and is given in appendix \ref{app:con}. Inserting $l_i$ into the action \re{ActionDTII}, we get the minimally coupled action
\bea \label{ActionDHOST}
\!\!\!\!\!\!\!\!\!\!\!\!\!\!\!\!\!\!\!\!\!\!\!\!\!\!\!\!\!\! 
  S & = & \int \rmd^4x \sqrt{-g} \bigg(\frac{1}{2} \mathring{R} - \frac{1}{2} \mathcal{G}_{2} K X - \mathcal{G}_{3} V + \mathcal{B}_{1} + \mathcal{B}_{2} Y + \mathcal{B}_{3} X^{\a\b} Y_{\a\b} \el
&& + \mathcal{A}_{1} Y_{\a\b} Y^{\a\b} + \mathcal{A}_{2} Y^2 + \mathcal{A}_{3} X^{\a\b} Y_{\a\b} Y + \mathcal{A}_{4} X^{\a\b} Y_{\a}{}^{\c} Y_{\b\c} + \mathcal{A}_{5} X^{\a\b} X^{\c\d} Y_{\a\b} Y_{\c\d}\bigg) \, ,
\eea
where the coefficients $\mathcal{B}_{i}(\varphi, X)$ and $\mathcal{A}_{i}(\varphi, X)$ are again relegated to appendix \ref{app:con}. If $\a_3=0$, then $\mathcal{B}_{i}=\mathcal{A}_{i}=0$, and \re{ActionDHOST} reduces to \re{actiona3zero}. The terms on the second line are non-Horndeski, but the functions $\mathcal{A}_{i}$ satisfy the conditions for the theory to be DHOST  \cite{Langlois:2018dxi, Kobayashi:2019hrl}. In order to obtain a minimally coupled action, it was important to consider coupling to $\dR_{\a\b}$, which vanishes for the Levi--Civita connection, rather than $\hat R_{\a\b}$.

To second order in $X_{\a\b}$, the action \re{ActionDHOST} reads
\bea \label{ActionDHOSTSO}
  S & = & \int \rmd^4x \sqrt{-g} \bigg\{\frac{1}{2} \mathring{R} - \frac{K X}{2 F} [ 1 - (\a_{1} + \a_{2} - \a_{3}) X ] \el
  && - \frac{V}{F^{2}} \Big[1 - \big(2 \a_{1} + \frac{1}{2} [ \a_{2} - \a_{3} ] \big) X + \big(\a_{1}^{2} + \frac{5}{8} \a_{2}^{2} + 2 \a_{1} [ \a_{2} - \a_{3} ] - \frac{3}{4} \a_{2} \a_{3} + \frac{1}{8}\a_{3}^{2}\big) X^{2}\Big] \el
  && + \frac{5}{8} \a_{3}^{2} X Y_{\a\b} Y^{\a\b} - \frac{13}{24} \a_{3}^{2} X Y^{2} - \frac{11}{12} \a_{3}^{2} X^{\a\b} Y_{\a\b} Y + \frac{5}{6} \a_{3}^{2} X^{\a\b} Y_{\a}{}^{\c} Y_{\b\c} \bigg\} \ .
\eea

In \cite{Aoki:2018lwx, Aoki:2019rvi} where the connection was unconstrained,  the couplings to $\mathring R_{\a\b}$ were instead eliminated by writing them in terms of the commutator of the Levi--Civita covariant derivative, without transforming to the Einstein frame. This leads to a different form of the action; it is well known that a Horndeski or a DHOST theory can take quite different-looking forms. Transforming the action in \cite{Aoki:2018lwx, Aoki:2019rvi} (keeping only the same original terms that we have) to the Einstein frame gives, to second order in $X_{\a\b}$,
\bea \label{AokiAction}
  S &=& \int \rmd^4x \sqrt{-g} \bigg\{\frac{1}{2} \mathring{R} - \frac{K X}{2 F} \big[ 1 - (\a_{1} + \a_{2} - \a_{3}) X \big] \el
  && - \frac{V}{F^{2}} \Big[ 1 - \big(2 \a_{1} + \frac{1}{2} [\a_{2} - \a_{3}] \big) X + \big( \a_{1}^{2} + \frac{5}{8} \a_{2}^{2} + 2 \a_{1} [\a_{2} - \a_{3}] - \frac{3}{4} \a_{2} \a_{3} + \frac{1}{8}\a_{3}^{2}\big) X^{2}\Big] \el
  && + \frac{1}{2} \a_{3}^{2} X Y_{\a\b} Y^{\a\b} - \frac{1}{2} \a_{3}^{2} X Y^{2} - \a_{3}^{2} X^{\a\b} Y_{\a\b} Y + \a_{3}^{2} X^{\a\b} Y_{\a}{}^{\c} Y_{\b\c}\bigg\} \ .
\eea
Comparing \re{ActionDHOSTSO} and \re{AokiAction} shows that the theories agree for $\a_3=0$, as then it makes no difference whether or not the torsion is constrained to be zero, as shown in \sec{sec:dis}. For $\a_3\neq0$ the theory with an unconstrained connection and the theory with zero torsion are physically inequivalent.

In \cite{Annala:2022gtl} it was shown that an action that depends on $\hat R_{\a\b}$ has a ghost around Minkowski space in the zero torsion case, and that in the unconstrained case there is a ghost around some FLRW backgrounds. This is not in contradiction with our result and the results of \cite{Aoki:2018lwx, Aoki:2019rvi} that these cases are stable. In \cite{Annala:2022gtl} it was assumed that the Legendre transformation to the Einstein frame is non-degenerate, which means that all degrees of freedom in $\hat R_{\a\b}$ are included in the Einstein frame action. In our case with a scalar field, there are no vector or tensor modes. In  \cite{Annala:2022gtl} the FLRW ghost was in the tensor sector.

\subsection{Metric case} \label{sec:comp}

Let us compare the Palatini case result \re{ActionDHOST} to the metric case. We are interested in the leading order differences in slow-roll inflation to see how the cases could be distinguished observationally. We again start with the action \re{action1}, now assuming that the connection is Levi--Civita. The action is of the Horndeski form when the kinetic term couples only to the Einstein tensor $G_{\a\b}=R_{\a\b}-\ha g_{\a\b}R$, not to the Ricci tensor and the Ricci scalar separately, otherwise it has a ghost. So we set $\a_1=-\ha \a_2$, $\a_3=0$. We again shift to the Einstein frame with the disformal transformation \re{disformal}. Now the calculation is more involved, because the Riemann tensor depends on the metric and its first and second derivative, unlike in the Palatini case. Hence, it is not invariant under the disformal transformation, which now introduces second order derivatives of $\varphi$. The transformation rules of the connection, the Riemann tensor, the Ricci tensor and the Ricci scalar are somewhat lengthy, and are given in appendix \ref{app:Riemann}. Inserting the result of the disformal transformation into the action \re{action1}, expanding to second order in $X_{\a\b}$ and choosing the disformal functions $\c_1$ and $\c_2$ so that the non-minimal couplings vanish, we get the Einstein frame action
\bea \label{actionmetric}
  S &=& \int \rmd^4 x \sqrt{-g} \Bigg\{ \ha \mathring R - \frac{K X}{2 F} \left(1 - \frac{\a_{2} X}{2}\right) - \frac{3}{4} \frac{F'{}^2}{F^{2}} X - \frac{V}{F^{2}} \left(1 + \frac{\a_{2} X}{2} - \frac{\a_{2}^{2} X^{2}}{8}\right) \el
  && + \frac{3}{2} \frac{(\a_{2} F)'}{2 F} \frac{F'}{F} X^2 - \frac{(\a_{2}F)'}{2 F} X Y + \frac{(\a_{2}F)'}{2 F} X^{\a\b} Y_{\a\b} \el
  && + \frac{1}{2} \a_{2}^{2} X^{\a\b} Y_{\a\b} Y - \frac{1}{2} \a_{2}^{2} X^{\a\b} Y_{\a}{}^{\c} Y_{\b\c}\Bigg\} \ .
\eea

To first order in $X_{\a\b}$ and $Y_{\a\b}$, we recover the result of the original New Higgs Inflation paper \cite{Germani:2010gm}, apart from the term involving $F'$. (The hybrid case with both $F'\neq0$ and $\a_2\neq0$ has been studied in \cite{Sato:2017qau}.) Apart from the $F'$ term, this leading order result agrees with the Palatini action \re{actiona3zero} when $\a_1=-\ha\a_2$, $\a_3=0$, as observed in \cite{Fumagalli:2020ody}. This is easy to understand: the distortion is sourced by $F'$ and $X_{\a\b}$, and only appears in the action via the total derivative and quadratic terms in the Riemann tensor \re{Riemann} and the coupling to the kinetic terms. So if $F'=0$, the distortion only enters at second order in $X_{\a\b}$. The second order terms on the first line of \re{actionmetric} also agree with the Palatini result, which is less obvious. It is only the non-Horndeski terms that are different.

However, in the Palatini case we can obtain the same action to first (but not second) order in $X_{\a\b}$ and $Y_{\a\b}$ by coupling to just $R$, \ie with $\a_2=\a_3=0$. In the metric case such a coupling would lead to a ghost. Also, in the metric case the derivatives of $F$ and $\a_2$ enter, unlike in the Palatini case. The terms involving $Y_{\a\b}$ are also different: in the Palatini case they appear only if $\a_3\neq0$. If the dynamics are dominated by the derivative coupling, the differences are small in slow-roll, but if the non-derivative coupling is important, the theories can have quite different predictions, as comparison of \cite{Sato:2017qau} and \cite{Gialamas:2020vto} shows.

\section{Conclusions} \label{sec:conc}

We have considered a theory where a scalar field kinetic term couples linearly to the Ricci tensor and the co-Ricci tensor, which appear linearly in the action, while the field itself can have non-linear non-minimal couplings. We look at both the Palatini formulation and the metric formulation. Extending previous Palatini work, we consider the case when either the non-metricity or the torsion is taken to vanish a priori. To establish the stability properties of the different cases and compare them side-by-side, we use a disformal transformation, followed by solving for the connection and inserting the solution back into the action. In this way we reduce the different cases to metric gravity with the Einstein--Hilbert action minimally coupled to matter.

We find that all the Palatini cases we consider are ghost-free: they are either in the Horndeski or DHOST class. If there is no coupling to the co-Ricci tensor, the Palatini result is independent of the assumptions about the connection. Otherwise, the case with unconstrained connection and the case with zero torsion are physically different. (If non-metricity is zero, the co-Ricci tensor vanishes.) We expand the actions up to second order in the scalar field kinetic term and compare the differences.

At leading order, the metric case and all the Palatini cases all agree with each other. However,  in the Palatini case a much wider range of couplings is stable, for example it is possible to simply couple the Ricci scalar to the trace of the kinetic term, simplifying the model. At second order, the Horndeski terms agree in the Palatini and metric cases, but the beyond Horndeski terms are different. The detailed form of the terms beyond the leading order might appear contrived if written in the metric formulation to begin with, but in the original Palatini formulation they are simple. The Palatini formulation can be seen as a selection principle to determine which complicated derivative couplings should appear in a metric formulation action.

Higgs inflation driven by derivative couplings in the metric formulation does not have a unitarity problem, unlike the metric formulation of the original Higgs inflation scenario with a non-derivative coupling to the Ricci scalar alone \cite{Germani:2014hqa, Escriva:2016cwl, Fumagalli:2017cdo, Fumagalli:2020ody}. The theory is however sensitive to loop corrections  \cite{Fumagalli:2020ody}. It would be interesting to see whether these features change in the derivative-driven or hybrid Palatini case. In the case with a non-derivative non-minimal coupling to the Einstein tensor alone, the unitarity problem is ameliorated in the Palatini formulation \cite{Bauer:2010, Shaposhnikov:2020fdv, Enckell:2020lvn, McDonald:2020lpz, Antoniadis:2021axu, Mikura:2021clt, Ito:2021ssc, Karananas:2022byw}.

It is an interesting question how to characterise the stability properties of theories in the Palatini formulation without reducing the theory a metric equivalent or calculating propagators. In \cite{Aoki:2019rvi} projective symmetry was used to show that a theory is U-degenerate, as the ghosts appear only in the unphysical projective mode. Projective invariance does not guarantee the absence of ghosts in general \cite{Annala:2022gtl}, only in particular cases \cite{Aoki:2019rvi, BeltranJimenez:2019acz, BeltranJimenez:2020sqf}. It would be interesting to understand better theories whose structure is tuned to the projective symmetry so that it makes them stable, and in particular whether projective symmetry (which has only a vectorial gauge mode) can prevent terms that would lead to tensor ghosts.

\acknowledgments

We thank Katsuki Aoki and Keigo Shimada for helpful correspondence. HBN acknowledges the Mathematica xAct package \cite{xAct} used in the calculations.

\appendix

\section{Details of the solution for the connection in the zero torsion case} \label{app:con}

We give here details of the connection calculation in \sec{sec:zerot} in the case when $\a_3\neq0$ and the torsion is zero. The general solution of the connection equation of motion \re{LEoM} in terms of the coefficients \re{LGuess} is
\bea \label{lSolutions}
&& l_{1} = \frac{\mathcal{G}_{1} \mathcal{G}_{1}' X^{2}}{1 + \mathcal{G}_{1} X} \el&&
l_{2} = - \frac{2 \mathcal{G}_{1}^{2} \mathcal{G}_{1}' X^{3}}{1 - \mathcal{G}_{1}^{2} X^{2}} \el&&
l_{3} = \frac{2 \mathcal{G}_{1}}{2 - \mathcal{G}_{1} X} \el&&
l_{4} = \frac{6 \mathcal{G}_{1} (1 - \mathcal{G}_{1} X)}{(2 - \mathcal{G}_{1} X)^{2}} \el&&
l_{5} = - \frac{12 \partial_{X}\mathcal{G}_{1} X + 9 \mathcal{G}_{1}^{3} X^{2} + 5 \mathcal{G}_{1}^{4} X^{3} - 2 \mathcal{G}_{1}^{2} X (19 - 6 \partial_{X}\mathcal{G}_{1} X^{2}) + 6 \mathcal{G}_{1} (2 - 5 \partial_{X}\mathcal{G}_{1} X^{2})}{3 (2 - \mathcal{G}_{1} X) (1 + \mathcal{G}_{1} X) (6 - 5 \mathcal{G}_{1} X)} \el&&
l_{6} = \frac{2 \mathcal{G}_{1} (1 - \mathcal{G}_{1} X)}{6 - 3 \mathcal{G}_{1} X} \el&&
l_{7} = \frac{4 \mathcal{G}_{1} \big[ 1 - \mathcal{G}_{1} X (1 - \mathcal{G}_{1} X)\big]}{3 (2 - \mathcal{G}_{1} X)^{2}} \el&&
l_{8} = - \frac{2}{3 (2 - \mathcal{G}_{1} X)^{2} (1 + \mathcal{G}_{1} X) (6 - 5 \mathcal{G}_{1} X)} \big[ 12 \partial_{X}\mathcal{G}_{1} X + 6 \mathcal{G}_{1}^{4} X^{3} - 5 \mathcal{G}_{1}^{5} X^{4} + 3 \mathcal{G}_{1}^{3} \partial_{X}\mathcal{G}_{1} X^{4}  \el&& + 6 \mathcal{G}_{1} (2 - \partial_{X}\mathcal{G}_{1} X^{2}) - \mathcal{G}_{1}^{2} (8 X + 6 \partial_{X}\mathcal{G}_{1} X^{3}) \big] \el&&
l_{9} = \frac{2 (\mathcal{G}_{1}' + \mathcal{G}_{1}^{2} \mathcal{G}_{1}' X^{2})}{1 - \mathcal{G}_{1}^{2} X^{2}} \el&&
l_{10} = \frac{125 \mathcal{G}_{1}^{2} + 84 \partial_{X}\mathcal{G}_{1}}{264 - 220 \mathcal{G}_{1} X} + \frac{4 \mathcal{G}_{1}^{2}}{(2 - \mathcal{G}_{1} X)^{2}} - \frac{25 \mathcal{G}_{1}^{2} + 4 \partial_{X}\mathcal{G}_{1}}{12 (2 - \mathcal{G}_{1} X)} - \frac{28 (\mathcal{G}_{1}^{2} - \partial_{X}\mathcal{G}_{1})}{33 (1 + \mathcal{G}_{1} X)} \el&&
l_{11} = \frac{2 \mathcal{G}_{1}^{2} (1 - 2 \mathcal{G}_{1} X)}{3 (2 - \mathcal{G}_{1} X)^{2}} \el&&
l_{12} = \partial_{X}\mathcal{G}_{1} + \frac{\mathcal{G}_{1}^{2}}{12 - 6 \mathcal{G}_{1} X} + \frac{5 (\mathcal{G}_{1}^{2} - \partial_{X}\mathcal{G}_{1})}{11 (1 + \mathcal{G}_{1} X)} - \frac{125 \mathcal{G}_{1}^{2} + 84 \partial_{X}\mathcal{G}_{1}}{66 (6 - 5 \mathcal{G}_{1} X)} \el&&
l_{13} = \frac{2}{(2 - \mathcal{G}_{1} X)^{2} (1 + \mathcal{G}_{1} X)^{2} (6 - 5 \mathcal{G}_{1} X)} \big\{ \mathcal{G}_{1}^{3} X \big[15 + \mathcal{G}_{1} X (19 - 20 \mathcal{G}_{1} X)\big] \el
  && + \partial_{X}\mathcal{G}_{1} (2 - \mathcal{G}_{1} X) ( 6 + 10 \mathcal{G}_{1} X - 5 \mathcal{G}_{1}^2 X^2 - \mathcal{G}_{1}^3 X^3 ) \big\} \el&&
l_{14} = 2 \partial_{X}\mathcal{G}_{1} + \frac{125 \mathcal{G}_{1}^{2} + 84 \partial_{X}\mathcal{G}_{1}}{792 - 660 \mathcal{G}_{1} X} + \frac{5 \mathcal{G}_{1}^{2} + 4 \partial_{X}\mathcal{G}_{1}}{24 - 12 \mathcal{G}_{1} X} - \frac{\mathcal{G}_{1}^{2}}{(2 - \mathcal{G}_{1} X)^{2}} - \frac{\mathcal{G}_{1}^{2} + \partial_{X}\mathcal{G}_{1}}{1 - \mathcal{G}_{1} X} \el&& + \frac{31 (\mathcal{G}_{1}^{2} - \partial_{X}\mathcal{G}_{1})}{33 (1 + \mathcal{G}_{1} X)} \el&&
l_{15} = - \frac{2 \mathcal{G}_{1}}{3 (2 - \mathcal{G}_{1} X)^{2} (1 - \mathcal{G}_{1} X) (1 + \mathcal{G}_{1} X)^{2} (6 - 5 \mathcal{G}_{1} X)} \big[12 \partial_{X}\mathcal{G}_{1} - 126 \mathcal{G}_{1} \partial_{X}\mathcal{G}_{1} X  + 10 \mathcal{G}_{1}^{6} X^{4} \el&& - 12 \mathcal{G}_{1}^{4} X^{2} (3 + 5 \partial_{X}\mathcal{G}_{1} X^{2}) - \mathcal{G}_{1}^{5} X^{3} (7 - 15 \partial_{X}\mathcal{G}_{1} X^{2}) + \mathcal{G}_{1}^{3} X (73 + 33 \partial_{X}\mathcal{G}_{1} X^{2}) \el&& - 2 \mathcal{G}_{1}^{2} (26 - 57 \partial_{X}\mathcal{G}_{1} X^{2}) \big] \ .
\eea
The coefficients of the final Einstein frame action \re{ActionDHOST} are
\bea \label{ActionDHOSTCoes}
& & \mathcal{B}_{1} = \frac{3 \mathcal{G}_{1}^{2} \mathcal{G}_{1}'^{2} X^{5}}{4 - 4 \mathcal{G}_{1}^{2} X^{2}} \el&&
\mathcal{B}_{2} = - \mathcal{G}_{1} \mathcal{G}_{1}' X^{2} \el&&
\mathcal{B}_{3} = \frac{\mathcal{G}_{1} \mathcal{G}_{1}' X \big[ 1 + \mathcal{G}_{1} X^{2} (2 \mathcal{G}_{1} + 3 \partial_{X}\mathcal{G}_{1} X)\big]}{1 - \mathcal{G}_{1}^{2} X^{2}} \el&&
\mathcal{A}_{1} = \frac{\mathcal{G}_{1}^{2} X (5 - 4 \mathcal{G}_{1} X)}{2 (2 - \mathcal{G}_{1} X)^{2}} \el&&
\mathcal{A}_{2} = - \frac{\mathcal{G}_{1}^{2} X (13 - 12 \mathcal{G}_{1} X + \mathcal{G}_{1}^2 X^2 )}{6 (2 - \mathcal{G}_{1} X)^{2}} \el&&
\mathcal{A}_{3} = - \frac{1}{3} \mathcal{G}_{1} \Big[6 \partial_{X}\mathcal{G}_{1} X + \mathcal{G}_{1}  \frac{11 - 12 \mathcal{G}_{1} X + 4 \mathcal{G}_{1}^2 X^2 }{(2 - \mathcal{G}_{1} X)^{2}}\Big] \el&&
\mathcal{A}_{4} = \frac{4}{(2 - \mathcal{G}_{1} X)^{2} (1 + \mathcal{G}_{1} X)^{2} (6 - 5 \mathcal{G}_{1} X)} \Big\{ 2 (\partial_{X}\mathcal{G}_{1})^{2} X^{2} + 6 \mathcal{G}_{1}^{5} X^{3} - 5 \mathcal{G}_{1}^{6} X^{4} \el
&& + 5 \mathcal{G}_{1}^{4} \partial_{X}\mathcal{G}_{1} X^{4} - \mathcal{G}_{1}^{3} \partial_{X}\mathcal{G}_{1} X^{3} (16 -\partial_{X}\mathcal{G}_{1} X^{2}) + \mathcal{G}_{1} \partial_{X}\mathcal{G}_{1} X (14 + 3 \partial_{X}\mathcal{G}_{1} X^{2}) \el
&& + \mathcal{G}_{1}^{2} \big[5 + \partial_{X}\mathcal{G}_{1} X^{2} (5 - 4 \partial_{X}\mathcal{G}_{1} X^{2})\big]\Big\} \el&&
\mathcal{A}_{5} = - \frac{1}{3 (2 - \mathcal{G}_{1} X)^{2} (1 - \mathcal{G}_{1} X) (1 + \mathcal{G}_{1} X)^{2} (6 - 5 \mathcal{G}_{1} X)} \Big\{ 24 (\partial_{X}\mathcal{G}_{1})^{2} X + 20 \mathcal{G}_{1}^{8} X^{5} \el&& + 12 \mathcal{G}_{1} \partial_{X}\mathcal{G}_{1} (2 + \partial_{X}\mathcal{G}_{1} X^{2}) + 12 \mathcal{G}_{1}^{2} \partial_{X}\mathcal{G}_{1} X (1 - 25 \partial_{X}\mathcal{G}_{1} X^{2}) -\mathcal{G}_{1}^{7} (64 X^{4} - 60 \partial_{X}\mathcal{G}_{1} X^{6}) \el&& + \mathcal{G}_{1}^{3} \big[2 - 48 \partial_{X}\mathcal{G}_{1} X^{2} (9 - 5 \partial_{X}\mathcal{G}_{1} X^{2})\big] + \mathcal{G}_{1}^{6} X^{3} \big[23 - 9 \partial_{X}\mathcal{G}_{1} X^{2} (28 - 5 \partial_{X}\mathcal{G}_{1} X^{2})\big] \el&& - 2 \mathcal{G}_{1}^{4} X \big[ 62 - 3 \partial_{X}\mathcal{G}_{1} X^{2} (61 + 25 \partial_{X}\mathcal{G}_{1} X^{2})\big] \el&&
+ \mathcal{G}_{1}^{5} X^{2} [ 125 + 3 \partial_{X}\mathcal{G}_{1} X^{2} ( 62 - 63 \partial_{X}\mathcal{G}_{1} X^{2})]\Big\} \ .
\eea

\section{Disformal transformation in the metric formulation} \label{app:Riemann}

Under the disformal transformation \re{tildeginv}, the Levi--Civita connection transforms as
\bea \label{contrans}
\mathring{\Gamma}^{\c}{}_{\a\b} & \rightarrow & \mathring{\Gamma}^{\c}{}_{\a\b} + \frac{- \gamma_{1}' g_{\a\b} + 2 \gamma_{2} Y_{\a\b}}{2 (\gamma_{1} + \gamma_{2} X)} X^{\c} - \frac{1}{2 \gamma_{1}} (\partial_{X}\gamma_{1} g_{\a\b} + \partial_{X}\gamma_{2} X_{\a\b}) \mathring{\nabla}^{\c}X + \frac{\gamma_{1}'}{2 \gamma_{1}} \delta_{(\a}{}^{\c}X_{\b)} \el&& + \frac{\partial_{X}\gamma_{1}}{2 \gamma_{1}} \delta_{(\a}{}^{\c}\mathring{\nabla}_{\b)}X + \frac{1}{2 \gamma_{1} (\gamma_{1} + \gamma_{2} X)} \Big\{ \partial_{X}\gamma_{1} \gamma_{2} g_{\a\b} X^{\c}{}_{\d} \mathring{\nabla}^{\d}X \el&& + \big[(-2 \gamma_{1}' \gamma_{2} + \gamma_{1} \gamma_{2}') X^{\c} + \gamma_{2} \partial_{X}\gamma_{2} X^{\c}{}_{\d} \mathring{\nabla}^{\d}X\big] X_{\a\b} \el
  && + 2 (- \partial_{X}\gamma_{1} \gamma_{2} + \gamma_{1} \partial_{X}\gamma_{2}) X_{(\a}{}^{\c}\mathring{\nabla}_{\b)}X\Big\} \ .
\eea
For the Riemann tensor we get
\bea \label{riemanntrans}
  \mathring{R}^{\a}{}_{\b\c\d} & \rightarrow & \mathring{R}^{\a}{}_{\b\c\d} + \frac{\partial_{X}\gamma_{1} g_{\b[\c}\mathring{\nabla}^{\a}\mathring{\nabla}_{\d]}X - \partial_{X}\gamma_{1} \delta_{[\c}{}^{\a}\mathring{\nabla}_{|\b|}\mathring{\nabla}_{\d]}X + \partial_{X}\gamma_{2} X_{\b[\c}\mathring{\nabla}^{\a}\mathring{\nabla}_{\d]}X}{\gamma_{1}} \el
  && + \frac{\gamma_{1}' g_{\b[\c}Y_{\d]}{}^{\a} - \gamma_{1}' \delta_{[\c}{}^{\a}Y_{|\b|\d]} + 2 \gamma_{2} (X^{\a}\mathring{\nabla}_{[\c}Y_{|\b|\d]} - Y_{\b[\c}Y_{\d]}{}^{\a})}{\gamma_{1} + \gamma_{2} X} \el
  && + \frac{1}{2 \gamma_{1} (\gamma_{1} + \gamma_{2} X)^{2}} \Big\{\big[-2 \gamma_{1}^{2} \c''_{1} + 2 \gamma_{1}'^{2} \gamma_{2} X + \gamma_{1} (3 \gamma_{1}'^{2} - 2 \c''_{1} \gamma_{2} X + \gamma_{1}' \gamma_{2}' X)\big] g_{\b[\c}X_{\d]}{}^{\a} \el
  && + \big[-2 \gamma_{1}^{2} \partial_{X}\gamma_{1}' + 2 \partial_{X}\gamma_{1} \gamma_{1}' \gamma_{2} X + \gamma_{1} (3 \partial_{X}\gamma_{1} \gamma_{1}' + \gamma_{1}' \gamma_{2} - 2 \partial_{X}\gamma_{1}' \gamma_{2} X + \gamma_{1}' \partial_{X}\gamma_{2} X)\big] \el&& g_{\b[\c}X^{\a}\mathring{\nabla}_{\d]}X\Big\} + \frac{1}{2 \gamma_{1}^{2}} \big[(\partial_{X}\gamma_{1})^{2} \mathring{\nabla}_{\e}X \mathring{\nabla}^{\e}X g_{\b[\c}\delta_{\d]}{}^{\a} - \partial_{X}\gamma_{1} \partial_{X}\gamma_{2} \mathring{\nabla}_{\e}X \mathring{\nabla}^{\e}X \delta_{[\c}{}^{\a}X_{|\b|\d]} \el&& - (3 (\partial_{X}\gamma_{1})^{2} - 2 \gamma_{1} \partial_{X}^{2}\gamma_{1}) (g_{\b[\c}\mathring{\nabla}_{\d]}X\mathring{\nabla}^{\a}X - \delta_{[\c}{}^{\a}\mathring{\nabla}_{\d]}X\mathring{\nabla}_{\b}X)\big] \el&& + \frac{1}{2 \gamma_{1} (\gamma_{1} + \gamma_{2} X)} \big[\gamma_{1}'^{2} X g_{\b[\c}\delta_{\d]}{}^{\a} + 2 \partial_{X}\gamma_{1} \gamma_{1}' X^{\e} \mathring{\nabla}_{\e}X g_{\b[\c}\delta_{\d]}{}^{\a} \el&& - 2 \partial_{X}\gamma_{1} \gamma_{2} X^{\e} \mathring{\nabla}_{\e}X g_{\b[\c}Y_{\d]}{}^{\a} + 2 \partial_{X}\gamma_{1} \gamma_{2} X^{\e} \mathring{\nabla}_{\e}X \delta_{[\c}{}^{\a}Y_{|\b|\d]} + (4 \gamma_{1}' \gamma_{2} - 2 \gamma_{1} \gamma_{2}') \el&& X_{\b[\c}Y_{\d]}{}^{\a} - 2 \gamma_{2} \partial_{X}\gamma_{2} X^{\e} \mathring{\nabla}_{\e}X X_{\b[\c}Y_{\d]}{}^{\a} + (2 \partial_{X}\gamma_{1} \gamma_{2} - 2 \gamma_{1} \partial_{X}\gamma_{2}) (X_{[\c}{}^{\a}\mathring{\nabla}_{|\b|}\mathring{\nabla}_{\d]}X \el&& - X_{[\c}Y_{|\b|\d]}\mathring{\nabla}^{\a}X + X_{[\c}Y_{\d]}{}^{\a}\mathring{\nabla}_{\b}X - X_{\b}Y_{[\c}{}^{\a}\mathring{\nabla}_{\d]}X) - 2 \partial_{X}\gamma_{1} \gamma_{2} g_{\b[\c}X^{\a\e}\mathring{\nabla}_{|\e|}\mathring{\nabla}_{\d]}X \el&& - 2 \gamma_{2} \partial_{X}\gamma_{2} X_{\b[\c}X^{\a\e}\mathring{\nabla}_{|\e|}\mathring{\nabla}_{\d]}X\big] + \frac{1}{(\gamma_{1} + \gamma_{2} X)^{2}} \Big\{(\gamma_{1} \gamma_{2}' - \gamma_{1}' \gamma_{2}) X_{[\c}{}^{\a}Y_{|\b|\d]} + \big[\gamma_{2} (\partial_{X}\gamma_{1} \el&& + \gamma_{2}) - \gamma_{1} \partial_{X}\gamma_{2}\big] X^{\a}Y_{\b[\c}\mathring{\nabla}_{\d]}X\Big\} - \frac{1}{2 \gamma_{1}^{2} (\gamma_{1} + \gamma_{2} X)} \bigg\{(\partial_{X}\gamma_{1})^{2} \gamma_{2} X_{\e\f} \mathring{\nabla}^{\e}X \mathring{\nabla}^{\f}X g_{\b[\c}\delta_{\d]}{}^{\a} \el&& + \partial_{X}\gamma_{1} (\partial_{X}\gamma_{1} \gamma_{2} - \gamma_{1} \partial_{X}\gamma_{2}) \mathring{\nabla}_{\e}X \mathring{\nabla}^{\e}X g_{\b[\c}X_{\d]}{}^{\a} + \big[2 \gamma_{1}^{2} \c''_{1} - \gamma_{1}'^{2} \gamma_{2} X - \gamma_{1} \el&& (3 \gamma_{1}'^{2} - 2 \c''_{1} \gamma_{2} X + \gamma_{1}' \gamma_{2}' X)\big] \delta_{[\c}{}^{\a}X_{|\b|\d]} + (2 \partial_{X}\gamma_{1} \gamma_{1}' \gamma_{2} + \gamma_{1} \gamma_{1}' \partial_{X}\gamma_{2} - \gamma_{1} \partial_{X}\gamma_{1} \gamma_{2}') \el&& X^{\e} \mathring{\nabla}_{\e}X \delta_{[\c}{}^{\a}X_{|\b|\d]} - \partial_{X}\gamma_{1} \gamma_{2} \partial_{X}\gamma_{2} X_{\e\f} \mathring{\nabla}^{\e}X \mathring{\nabla}^{\f}X \delta_{[\c}{}^{\a}X_{|\b|\d]} - \big[2 \gamma_{1}^{2} \partial_{X}\gamma_{1}' \el&& - 2 \partial_{X}\gamma_{1} \gamma_{1}' \gamma_{2} X - \gamma_{1} (3 \partial_{X}\gamma_{1} \gamma_{1}' - 2 \partial_{X}\gamma_{1}' \gamma_{2} X + \gamma_{1}' \partial_{X}\gamma_{2} X)\big] g_{\b[\c}X_{\d]}\mathring{\nabla}^{\a}X \el&& + \big[2 \gamma_{1}^{2} \partial_{X}\gamma_{1}' - 2 \partial_{X}\gamma_{1} \gamma_{1}' \gamma_{2} X - \gamma_{1} (3 \partial_{X}\gamma_{1} \gamma_{1}' - 2 \partial_{X}\gamma_{1}' \gamma_{2} X + \gamma_{1}' \partial_{X}\gamma_{2} X)\big] (\delta_{[\c}{}^{\a}X_{\d]}\mathring{\nabla}_{\b}X \el&& + \delta_{[\c}{}^{\a}X_{|\b|}\mathring{\nabla}_{\d]}X) - \Big\{(\partial_{X}\gamma_{1})^{2} \gamma_{2} - 2 \partial_{X}\gamma_{1} \partial_{X}\gamma_{2} (2 \gamma_{1} + \gamma_{2} X) + \gamma_{1} \big[2 \gamma_{1} \partial_{X}^{2}\gamma_{2} - ((\partial_{X}\gamma_{2})^{2} \el&& - 2 \gamma_{2} \partial_{X}^{2}\gamma_{2}) X\big]\Big\} X_{\b[\c}\mathring{\nabla}_{\d]}X\mathring{\nabla}^{\a}X - \partial_{X}\gamma_{1} (\partial_{X}\gamma_{1} \gamma_{2} - \gamma_{1} \partial_{X}\gamma_{2}) g_{\b[\c}X_{\d]}{}^{\e}\mathring{\nabla}^{\a}X\mathring{\nabla}_{\e}X \el&& + \partial_{X}\gamma_{1} (\partial_{X}\gamma_{1} \gamma_{2} - \gamma_{1} \partial_{X}\gamma_{2}) \delta_{[\c}{}^{\a}X_{\d]}{}^{\e}\mathring{\nabla}_{\b}X\mathring{\nabla}_{\e}X + \partial_{X}\gamma_{1} (\partial_{X}\gamma_{1} \gamma_{2} - \gamma_{1} \partial_{X}\gamma_{2}) \el&& \delta_{[\c}{}^{\a}X_{|\b|}{}^{\e}\mathring{\nabla}_{\d]}X\mathring{\nabla}_{\e}X\bigg\} + \frac{1}{2 \gamma_{1}^{2} (\gamma_{1} + \gamma_{2} X)^{2}} \bigg\{\big[\gamma_{1} (3 \partial_{X}\gamma_{1} \gamma_{1}' \gamma_{2} - 2 \gamma_{1} \partial_{X}\gamma_{1}' \gamma_{2} + \gamma_{1} \gamma_{1}' \partial_{X}\gamma_{2} \el&& - \gamma_{1} \partial_{X}\gamma_{1} \gamma_{2}') + \gamma_{2} (2 \partial_{X}\gamma_{1} \gamma_{1}' \gamma_{2} - 2 \gamma_{1} \partial_{X}\gamma_{1}' \gamma_{2} + \gamma_{1} \gamma_{1}' \partial_{X}\gamma_{2}) X\big] X^{\e} \mathring{\nabla}_{\e}X g_{\b[\c}X_{\d]}{}^{\a} \el&& + \gamma_{2} \big[\gamma_{1}^{2} (2 \partial_{X}\gamma_{1}' + \gamma_{2}') - 2 \partial_{X}\gamma_{1} \gamma_{1}' \gamma_{2} X - \gamma_{1} (3 \partial_{X}\gamma_{1} \gamma_{1}' + 2 \gamma_{1}' \gamma_{2} - 2 \partial_{X}\gamma_{1}' \gamma_{2} X \el&& + \gamma_{1}' \partial_{X}\gamma_{2} X)\big] X_{\b[\c}X^{\a}\mathring{\nabla}_{\d]}X - \Big\{2 \gamma_{1}^{3} \partial_{X}^{2}\gamma_{2} + 3 (\partial_{X}\gamma_{1})^{2} \gamma_{2}^{2} X - \gamma_{1}^{2} (2 \partial_{X}^{2}\gamma_{1} \gamma_{2} \el&& + 4 \partial_{X}\gamma_{1} \partial_{X}\gamma_{2} + \gamma_{2} \partial_{X}\gamma_{2} + (\partial_{X}\gamma_{2})^{2} X - 2 \gamma_{2} \partial_{X}^{2}\gamma_{2} X) + \gamma_{1} \gamma_{2} \big[4 (\partial_{X}\gamma_{1})^{2} \el&& - 2 \partial_{X}^{2}\gamma_{1} \gamma_{2} X + \partial_{X}\gamma_{1} (\gamma_{2} - 2 \partial_{X}\gamma_{2} X)\big]\Big\} X_{[\c}{}^{\a}\mathring{\nabla}_{\d]}X\mathring{\nabla}_{\b}X + \big[\gamma_{1}^{2} (2 \partial_{X}^{2}\gamma_{1} \gamma_{2} \el&& + \partial_{X}\gamma_{1} \partial_{X}\gamma_{2}) - 3 (\partial_{X}\gamma_{1})^{2} \gamma_{2}^{2} X - \gamma_{1} \gamma_{2} (4 (\partial_{X}\gamma_{1})^{2} + \partial_{X}\gamma_{1} \gamma_{2} - 2 \partial_{X}^{2}\gamma_{1} \gamma_{2} X)\big] \el&& g_{\b[\c}X^{\a\e}\mathring{\nabla}_{\d]}X\mathring{\nabla}_{\e}X - \gamma_{2} \big[2 \gamma_{1}^{2} \partial_{X}^{2}\gamma_{2} - 2 \partial_{X}\gamma_{1} \gamma_{2} \partial_{X}\gamma_{2} X - \gamma_{1} (3 \partial_{X}\gamma_{1} \partial_{X}\gamma_{2} \el&& + \gamma_{2} \partial_{X}\gamma_{2} + (\partial_{X}\gamma_{2})^{2} X - 2 \gamma_{2} \partial_{X}^{2}\gamma_{2} X)\big] X_{\b[\c}X^{\a\e}\mathring{\nabla}_{\d]}X\mathring{\nabla}_{\e}X\bigg\} \ .
\eea
The Ricci tensor transforms as
\bea \label{riccitensortrans}
  \mathring{R}_{\a\b} & \rightarrow & \mathring{R}_{\a\b} + \frac{-2 \gamma_{1} \gamma_{1}' - 3 \gamma_{1}' \gamma_{2} X + \gamma_{1} \gamma_{2}' X}{2 (\gamma_{1} + \gamma_{2} X)^{2}} Y_{\a\b} + \frac{- \gamma_{1}' g_{\a\b} Y + 2 \gamma_{2} (Y_{\a\b} Y + X^{\c} \mathring{\nabla}_{\c}Y_{\a\b})}{2 (\gamma_{1} + \gamma_{2} X)} \el
  && - \frac{(\partial_{X}\gamma_{1} g_{\a\b} + \partial_{X}\gamma_{2} X_{\a\b}) \mathring{\nabla}_{\c}\mathring{\nabla}^{\c}X}{2 \gamma_{1}} - \frac{1}{4 \gamma_{1} (\gamma_{1} + \gamma_{2} X)^{2}} \bigg\{X \big[2 \gamma_{1}^{2} \c''_{1} + \gamma_{1}'^{2} \gamma_{2} X \el
  && + \gamma_{1} (2 \c''_{1} \gamma_{2} - \gamma_{1}' \gamma_{2}') X\big] g_{\a\b} + \Big\{ 4 \gamma_{1}^{2} \partial_{X}\gamma_{1}' + 2 \partial_{X}\gamma_{1} \gamma_{1}' \gamma_{2} X + \gamma_{1} \big[ 4 \partial_{X}\gamma_{1}' \gamma_{2} X \el
  && - \partial_{X}\gamma_{1} \gamma_{2}' X - \gamma_{1}' (\gamma_{2} + \partial_{X}\gamma_{2} X)\big]\Big\} g_{\a\b} X^{\c} \mathring{\nabla}_{\c}X - 2 \big[2 \gamma_{1}^{2} \partial_{X}\gamma_{2} + \partial_{X}\gamma_{1} \gamma_{2}^{2} X \el
  && + \gamma_{1} \gamma_{2} (- \gamma_{2} + \partial_{X}\gamma_{2} X)\big] X^{\c} Y_{\a\b} \mathring{\nabla}_{\c}X\bigg\} - \frac{1}{4 \gamma_{1}^{2} (\gamma_{1} + \gamma_{2} X)} \Big\{\big[2 \gamma_{1}^{2} \partial_{X}^{2}\gamma_{1} - (\partial_{X}\gamma_{1})^{2} \gamma_{2} X \el
  && + \gamma_{1} (2 \partial_{X}^{2}\gamma_{1} \gamma_{2} + \partial_{X}\gamma_{1} \partial_{X}\gamma_{2}) X\big] g_{\a\b} \mathring{\nabla}_{\c}X \mathring{\nabla}^{\c}X + \big[2 (\partial_{X}\gamma_{1})^{2} \gamma_{2} + \partial_{X}\gamma_{1} (-2 \gamma_{1} \partial_{X}\gamma_{2} \el&& + \gamma_{2} \partial_{X}\gamma_{2} X) + \gamma_{1} (2 \gamma_{1} \partial_{X}^{2}\gamma_{2} - (\partial_{X}\gamma_{2})^{2} X + 2 \gamma_{2} \partial_{X}^{2}\gamma_{2} X)\big] X_{\a\b} \mathring{\nabla}_{\c}X \mathring{\nabla}^{\c}X\Big\} \el&& + \frac{1}{4 \gamma_{1}^{2} (\gamma_{1} + \gamma_{2} X)^{2}} \bigg\{\big[-4 \gamma_{1}^{3} \c''_{1} + 3 \gamma_{1}'^{2} \gamma_{2}^{2} X^{2} + 2 \gamma_{1}^{2} (3 \gamma_{1}'^{2} - 5 \c''_{1} \gamma_{2} X + \gamma_{1}' \gamma_{2}' X) \el&& + \gamma_{1} \gamma_{2} X (10 \gamma_{1}'^{2} - 6 \c''_{1} \gamma_{2} X + 3 \gamma_{1}' \gamma_{2}' X)\big] X_{\a\b} + 2 \big[- \gamma_{1}^{3} (4 \partial_{X}\gamma_{1}' + \gamma_{2}') \el&& + 4 \partial_{X}\gamma_{1} \gamma_{1}' \gamma_{2}^{2} X^{2} + 2 \gamma_{1} \gamma_{2} X (5 \partial_{X}\gamma_{1} \gamma_{1}' - 2 \partial_{X}\gamma_{1}' \gamma_{2} X + \gamma_{1}' \partial_{X}\gamma_{2} X) + \gamma_{1}^{2} (6 \partial_{X}\gamma_{1} \gamma_{1}' \el&& + \gamma_{1}' \gamma_{2} - 8 \partial_{X}\gamma_{1}' \gamma_{2} X + 2 \gamma_{1}' \partial_{X}\gamma_{2} X)\big] X_{(\a} \mathring{\nabla}_{\b)}X + \Big\{3 (\partial_{X}\gamma_{1})^{2} \gamma_{2}^{2} X^{2} + 2 \gamma_{1} \gamma_{2} X (4 (\partial_{X}\gamma_{1})^{2} \el&& - \partial_{X}^{2}\gamma_{1} \gamma_{2} X + \partial_{X}\gamma_{1} \partial_{X}\gamma_{2} X) - 2 \gamma_{1}^{3} (2 \partial_{X}^{2}\gamma_{1} + \partial_{X}\gamma_{2} + \partial_{X}^{2}\gamma_{2} X) + \gamma_{1}^{2} \big[6 (\partial_{X}\gamma_{1})^{2} + \gamma_{2}^{2} \el&& + (\partial_{X}\gamma_{2})^{2} X^{2} + 2 \partial_{X}\gamma_{1} (\gamma_{2} + 2 \partial_{X}\gamma_{2} X) - 2 \gamma_{2} X (3 \partial_{X}^{2}\gamma_{1} + \partial_{X}^{2}\gamma_{2} X)\big]\Big\} \mathring{\nabla}_{\a}X \mathring{\nabla}_{\b}X \el&& + \big[- \gamma_{1}^{2} (4 \partial_{X}\gamma_{1}' \gamma_{2} + 2 \gamma_{1}' \partial_{X}\gamma_{2} - 2 \partial_{X}\gamma_{1} \gamma_{2}' + \gamma_{2} \gamma_{2}') - 2 \partial_{X}\gamma_{1} \gamma_{1}' \gamma_{2}^{2} X + \gamma_{1} \gamma_{2} (2 \gamma_{1}' \gamma_{2} \el&& - 4 \partial_{X}\gamma_{1}' \gamma_{2} X - \gamma_{1}' \partial_{X}\gamma_{2} X + 3 \partial_{X}\gamma_{1} \gamma_{2}' X)\big] X_{\a\b} X^{\c} \mathring{\nabla}_{\c}X + 2 \big[2 \gamma_{1}^{3} \partial_{X}^{2}\gamma_{2} + (\partial_{X}\gamma_{1})^{2} \gamma_{2}^{2} X \el&& + \gamma_{1} \gamma_{2} (2 (\partial_{X}\gamma_{1})^{2} + \partial_{X}\gamma_{1} \gamma_{2} - 2 \partial_{X}^{2}\gamma_{1} \gamma_{2} X) - \gamma_{1}^{2} (2 \partial_{X}^{2}\gamma_{1} \gamma_{2} + 2 \partial_{X}\gamma_{1} \partial_{X}\gamma_{2} + \gamma_{2} \partial_{X}\gamma_{2} \el&& + (\partial_{X}\gamma_{2})^{2} X - 2 \gamma_{2} \partial_{X}^{2}\gamma_{2} X)\big] X_{(\a|\c|} \mathring{\nabla}_{\b)}X \mathring{\nabla}^{\c}X + \Big\{\gamma_{1} \big[2 \gamma_{1} \partial_{X}^{2}\gamma_{1} \gamma_{2} - \partial_{X}\gamma_{1} \gamma_{2} (2 \partial_{X}\gamma_{1} + \gamma_{2}) \el&& + 2 \gamma_{1} \partial_{X}\gamma_{1} \partial_{X}\gamma_{2}\big] + \gamma_{2} (- (\partial_{X}\gamma_{1})^{2} \gamma_{2} + 2 \gamma_{1} \partial_{X}^{2}\gamma_{1} \gamma_{2} + \gamma_{1} \partial_{X}\gamma_{1} \partial_{X}\gamma_{2}) X\Big\} g_{\a\b} X_{\c\d} \mathring{\nabla}^{\c}X \mathring{\nabla}^{\d}X \el
  && + \gamma_{2} \Big\{2 \gamma_{1}^{2} \partial_{X}^{2}\gamma_{2} + \partial_{X}\gamma_{1} \gamma_{2} \partial_{X}\gamma_{2} X - \gamma_{1} \big[(\partial_{X}\gamma_{2})^{2} X + \gamma_{2} (\partial_{X}\gamma_{2} - 2 \partial_{X}^{2}\gamma_{2} X)\big]\Big\} \el
  && X_{\a\b} X_{\c\d} \mathring{\nabla}^{\c}X \mathring{\nabla}^{\d}X\bigg\} - \frac{1}{2 \gamma_{1} (\gamma_{1} + \gamma_{2} X)} \Big\{(2 \gamma_{1}' \gamma_{2} - \gamma_{1} \gamma_{2}') X_{\a\b} Y + 2 (\partial_{X}\gamma_{1} \gamma_{2} - \gamma_{1} \partial_{X}\gamma_{2}) \el
  && X_{(\a} Y \mathring{\nabla}_{\b)}X + \big[\partial_{X}\gamma_{1} \gamma_{2} X + \gamma_{1} (2 \partial_{X}\gamma_{1} + \gamma_{2} + \partial_{X}\gamma_{2} X)\big] \mathring{\nabla}_{\a}\mathring{\nabla}_{\b}X \el
  && - \partial_{X}\gamma_{1} \gamma_{2} g_{\a\b} X^{\c} Y \mathring{\nabla}_{\c}X - \gamma_{2} \partial_{X}\gamma_{2} X_{\a\b} X^{\c} Y \mathring{\nabla}_{\c}X + 2 (- \partial_{X}\gamma_{1} \gamma_{2} + \gamma_{1} \partial_{X}\gamma_{2}) X_{(\a} Y_{\b)\c} \mathring{\nabla}^{\c}X \el
  && + 2 (\partial_{X}\gamma_{1} \gamma_{2} - \gamma_{1} \partial_{X}\gamma_{2}) X_{(\a|\c|} \mathring{\nabla}^{\c}\mathring{\nabla}_{\b)}X - \partial_{X}\gamma_{1} \gamma_{2} g_{\a\b} X_{\c\d} \mathring{\nabla}^{\d}\mathring{\nabla}^{\c}X \el
  && - \gamma_{2} \partial_{X}\gamma_{2} X_{\a\b} X_{\c\d} \mathring{\nabla}^{\d}\mathring{\nabla}^{\c}X\Big\} \ .
\eea
Finally, the Ricci scalar transforms as
\bea \label{ricciscalartrans}
  \mathring{R} & \rightarrow & \frac{\mathring{R}}{\gamma_{1}} + \frac{1}{2 \gamma_{1} (\gamma_{1} + \gamma_{2} X)^{2}} \Big\{3 X \big[\gamma_{1}'^{2} + \gamma_{1}' \gamma_{2}' X - 2 \partial^{2}_{\varphi}\gamma_{1} (\gamma_{1} + \gamma_{2} X)\big] \el
  && - 2 (3 \gamma_{1} \gamma_{1}' + 4 \gamma_{1}' \gamma_{2} X - \gamma_{1} \gamma_{2}' X) Y + \big[6 \partial_{X}\gamma_{1} \gamma_{1}' + 4 \gamma_{1}' \gamma_{2} - \gamma_{1} (12 \partial_{X}\gamma_{1}' + \gamma_{2}') \el
  && - 12 \partial_{X}\gamma_{1}' \gamma_{2} X + 3 \gamma_{1}' \partial_{X}\gamma_{2} X + 3 \partial_{X}\gamma_{1} \gamma_{2}' X\big] X^{\a} \mathring{\nabla}_{\a}X\Big\} \el
  && + \frac{1}{2 \gamma_{1}^{2} (\gamma_{1} + \gamma_{2} X)^{2}} \Big\{\big[2 \gamma_{1}^{2} \partial_{X}\gamma_{2} + \partial_{X}\gamma_{1} \gamma_{2}^{2} X - \gamma_{1} \gamma_{2} (\gamma_{2} - \partial_{X}\gamma_{2} X)\big] X^{\a} Y \mathring{\nabla}_{\a}X \el
  && + \big[2 \gamma_{1}^{2} \partial_{X}\gamma_{2} + 3 \partial_{X}\gamma_{1} \gamma_{2}^{2} X + \gamma_{1} \gamma_{2} (2 \partial_{X}\gamma_{1} - \gamma_{2} + \partial_{X}\gamma_{2} X)\big] \mathring{\nabla}^{\a}X \mathring{\nabla}_{\b}X_{\a}{}^{\b}\Big\} \el
  && - \frac{\gamma_{2} (\mathring{R}^{\a\b} X_{\a\b} - \mathring{\nabla}_{\b}\mathring{\nabla}_{\a}X^{\a\b})}{\gamma_{1} (\gamma_{1} + \gamma_{2} X)} \el
  && - \frac{1}{4 \gamma_{1}^{3} (\gamma_{1} + \gamma_{2} X)^{2}} \mathring{\nabla}^{\a}X \Big\{ \Big[-6 (\partial_{X}\gamma_{1})^{2} \gamma_{2}^{2} X^{2} + \gamma_{1} \gamma_{2} X (-10 (\partial_{X}\gamma_{1})^{2} + 3 \partial_{X}\gamma_{1} \gamma_{2} \el
  && + 8 \partial^{2}_{X}\gamma_{1} \gamma_{2} X + 2 \partial_{X}\gamma_{1} \partial_{X}\gamma_{2} X) + 2 \gamma_{1}^{3} (6 \partial^{2}_{X}\gamma_{1} + 3 \partial_{X}\gamma_{2} + 2 \partial^{2}_{X}\gamma_{2} X) + \gamma_{1}^{2} \big[-6 (\partial_{X}\gamma_{1})^{2} \el
  && - 3 \gamma_{2}^{2} - 2 (\partial_{X}\gamma_{2})^{2} X^{2} - 2 \partial_{X}\gamma_{1} (\gamma_{2} + \partial_{X}\gamma_{2} X) + \gamma_{2} X (20 \partial^{2}_{X}\gamma_{1} + \partial_{X}\gamma_{2} + 4 \partial^{2}_{X}\gamma_{2} X)\big]\Big] \mathring{\nabla}_{\a}X \el
  && + 2 \big[-2 \gamma_{1}^{3} \partial^{2}_{X}\gamma_{2} + 3 (\partial_{X}\gamma_{1})^{2} \gamma_{2}^{2} X + \gamma_{1} \gamma_{2} (5 (\partial_{X}\gamma_{1})^{2} + 2 \partial_{X}\gamma_{1} \gamma_{2} - 4 \partial^{2}_{X}\gamma_{1} \gamma_{2} X \el
  && - \partial_{X}\gamma_{1} \partial_{X}\gamma_{2} X) + \gamma_{1}^{2} (-4 \partial^{2}_{X}\gamma_{1} \gamma_{2} - 2 \partial_{X}\gamma_{1} \partial_{X}\gamma_{2} + \gamma_{2} \partial_{X}\gamma_{2} + (\partial_{X}\gamma_{2})^{2} X - 2 \gamma_{2} \partial^{2}_{X}\gamma_{2} X)\big] \el
  && X_{\a\b} \mathring{\nabla}^{\b}X \Big\} - \frac{1}{\gamma_{1}^{2} (\gamma_{1} + \gamma_{2} X)} \Big\{\big[2 \partial_{X}\gamma_{1} \gamma_{2} X + \gamma_{1} (3 \partial_{X}\gamma_{1} + \gamma_{2} + \partial_{X}\gamma_{2} X)\big] \mathring{\nabla}_{\a}\mathring{\nabla}^{\a}X \el
  && - (2 \partial_{X}\gamma_{1} \gamma_{2} + \gamma_{1} \partial_{X}\gamma_{2}) X_{\a\b} \mathring{\nabla}^{\b}\mathring{\nabla}^{\a}X\Big\} \ .
\eea

\bibliographystyle{JHEP}
\bibliography{new}

\providecommand{\href}[2]{#2}\begingroup\raggedright\begin{thebibliography}{100}

\bibitem{Starobinsky:1979ty}
A.~A. Starobinsky, \emph{{Spectrum of relict gravitational radiation and the
  early state of the universe}}, {\emph{JETP Lett.} {\bfseries 30} (1979)
  682--685}.

\bibitem{Starobinsky:1980te}
A.~A. Starobinsky, \emph{{A New Type of Isotropic Cosmological Models Without
  Singularity}},
  \href{http://dx.doi.org/10.1016/0370-2693(80)90670-X}{\emph{Phys. Lett.}
  {\bfseries 91B} (1980) 99--102}.

\bibitem{Kazanas:1980tx}
D.~Kazanas, \emph{{Dynamics of the Universe and Spontaneous Symmetry
  Breaking}}, \href{http://dx.doi.org/10.1086/183361}{\emph{Astrophys. J.}
  {\bfseries 241} (1980) L59--L63}.

\bibitem{Guth:1980zm}
A.~H. Guth, \emph{{The Inflationary Universe: A Possible Solution to the
  Horizon and Flatness Problems}},
  \href{http://dx.doi.org/10.1103/PhysRevD.23.347}{\emph{Phys. Rev.} {\bfseries
  D23} (1981) 347--356}.

\bibitem{Sato:1980yn}
K.~Sato, \emph{{First Order Phase Transition of a Vacuum and Expansion of the
  Universe}}, {\emph{Mon. Not. Roy. Astron. Soc.} {\bfseries 195} (1981)
  467--479}.

\bibitem{Mukhanov:1981xt}
V.~F. Mukhanov and G.~V. Chibisov, \emph{{Quantum Fluctuations and a
  Nonsingular Universe}}, {\emph{JETP Lett.} {\bfseries 33} (1981) 532--535}.

\bibitem{Linde:1981mu}
A.~D. Linde, \emph{{A New Inflationary Universe Scenario: A Possible Solution
  of the Horizon, Flatness, Homogeneity, Isotropy and Primordial Monopole
  Problems}}, \href{http://dx.doi.org/10.1016/0370-2693(82)91219-9}{\emph{Phys.
  Lett.} {\bfseries 108B} (1982) 389--393}.

\bibitem{Albrecht:1982wi}
A.~Albrecht and P.~J. Steinhardt, \emph{{Cosmology for Grand Unified Theories
  with Radiatively Induced Symmetry Breaking}},
  \href{http://dx.doi.org/10.1103/PhysRevLett.48.1220}{\emph{Phys. Rev. Lett.}
  {\bfseries 48} (1982) 1220--1223}.

\bibitem{Hawking:1981fz}
S.~W. Hawking and I.~G. Moss, \emph{{Supercooled Phase Transitions in the Very
  Early Universe}},
  \href{http://dx.doi.org/10.1016/0370-2693(82)90946-7}{\emph{Phys. Lett.}
  {\bfseries 110B} (1982) 35--38}.

\bibitem{Chibisov:1982nx}
G.~V. Chibisov and V.~F. Mukhanov, \emph{{Galaxy formation and phonons}},
  {\emph{Mon. Not. Roy. Astron. Soc.} {\bfseries 200} (1982) 535--550}.

\bibitem{Hawking:1982cz}
S.~W. Hawking, \emph{{The Development of Irregularities in a Single Bubble
  Inflationary Universe}},
  \href{http://dx.doi.org/10.1016/0370-2693(82)90373-2}{\emph{Phys. Lett.}
  {\bfseries 115B} (1982) 295}.

\bibitem{Guth:1982ec}
A.~H. Guth and S.~Y. Pi, \emph{{Fluctuations in the New Inflationary
  Universe}}, \href{http://dx.doi.org/10.1103/PhysRevLett.49.1110}{\emph{Phys.
  Rev. Lett.} {\bfseries 49} (1982) 1110--1113}.

\bibitem{Starobinsky:1982ee}
A.~A. Starobinsky, \emph{{Dynamics of Phase Transition in the New Inflationary
  Universe Scenario and Generation of Perturbations}},
  \href{http://dx.doi.org/10.1016/0370-2693(82)90541-X}{\emph{Phys. Lett.}
  {\bfseries 117B} (1982) 175--178}.

\bibitem{Sasaki:1986hm}
M.~Sasaki, \emph{{Large Scale Quantum Fluctuations in the Inflationary
  Universe}}, \href{http://dx.doi.org/10.1143/PTP.76.1036}{\emph{Prog. Theor.
  Phys.} {\bfseries 76} (1986) 1036}.

\bibitem{Mukhanov:1988jd}
V.~F. Mukhanov, \emph{{Quantum Theory of Gauge Invariant Cosmological
  Perturbations}}, {\emph{Sov. Phys. JETP} {\bfseries 67} (1988) 1297--1302}.

\bibitem{Planck:2018jri}
{\scshape Planck} collaboration, Y.~Akrami et~al., \emph{{Planck 2018 results.
  X. Constraints on inflation}},
  \href{http://dx.doi.org/10.1051/0004-6361/201833887}{\emph{Astron.
  Astrophys.} {\bfseries 641} (2020) A10},
  [\href{https://arxiv.org/abs/1807.06211}{{\ttfamily 1807.06211}}].

\bibitem{Callan:1970ze}
C.~G. Callan, Jr., S.~R. Coleman and R.~Jackiw, \emph{{A New improved energy -
  momentum tensor}},
  \href{http://dx.doi.org/10.1016/0003-4916(70)90394-5}{\emph{Annals Phys.}
  {\bfseries 59} (1970) 42--73}.

\bibitem{Bezrukov:2007}
F.~L. Bezrukov and M.~Shaposhnikov, \emph{{The Standard Model Higgs boson as
  the inflaton}},
  \href{http://dx.doi.org/10.1016/j.physletb.2007.11.072}{\emph{Phys. Lett.}
  {\bfseries B659} (2008) 703--706},
  [\href{https://arxiv.org/abs/0710.3755}{{\ttfamily 0710.3755}}].

\bibitem{Bezrukov:2013}
F.~Bezrukov, \emph{{The Higgs field as an inflaton}},
  \href{http://dx.doi.org/10.1088/0264-9381/30/21/214001}{\emph{Class. Quant.
  Grav.} {\bfseries 30} (2013) 214001},
  [\href{https://arxiv.org/abs/1307.0708}{{\ttfamily 1307.0708}}].

\bibitem{Bezrukov:2015}
F.~Bezrukov and M.~Shaposhnikov, \emph{{Inflation, LHC and the Higgs boson}},
  \href{http://dx.doi.org/10.1016/j.crhy.2015.08.005}{\emph{Comptes Rendus
  Physique} {\bfseries 16} (2015) 994--1002}.

\bibitem{Rubio:2018ogq}
J.~Rubio, \emph{{Higgs inflation}},
  \href{http://dx.doi.org/10.3389/fspas.2018.00050}{\emph{Front. Astron. Space
  Sci.} {\bfseries 5} (2019) 50},
  [\href{https://arxiv.org/abs/1807.02376}{{\ttfamily 1807.02376}}].

\bibitem{Capozziello:1999uwa}
S.~Capozziello and G.~Lambiase, \emph{{Nonminimal derivative coupling and the
  recovering of cosmological constant}},
  \href{http://dx.doi.org/10.1023/A:1026631531309}{\emph{Gen. Rel. Grav.}
  {\bfseries 31} (1999) 1005--1014},
  [\href{https://arxiv.org/abs/gr-qc/9901051}{{\ttfamily gr-qc/9901051}}].

\bibitem{Capozziello:1999xt}
S.~Capozziello, G.~Lambiase and H.~J. Schmidt, \emph{{Nonminimal derivative
  couplings and inflation in generalized theories of gravity}},
  \href{http://dx.doi.org/10.1002/(SICI)1521-3889(200001)9:1<39::AID-ANDP39>3.0.CO}{\emph{Annalen
  Phys.} {\bfseries 9} (2000) 39--48},
  [\href{https://arxiv.org/abs/gr-qc/9906051}{{\ttfamily gr-qc/9906051}}].

\bibitem{Daniel:2007kk}
S.~F. Daniel and R.~R. Caldwell, \emph{{Consequences of a cosmic scalar with
  kinetic coupling to curvature}},
  \href{http://dx.doi.org/10.1088/0264-9381/24/22/017}{\emph{Class. Quant.
  Grav.} {\bfseries 24} (2007) 5573--5580},
  [\href{https://arxiv.org/abs/0709.0009}{{\ttfamily 0709.0009}}].

\bibitem{Sushkov:2009hk}
S.~V. Sushkov, \emph{{Exact cosmological solutions with nonminimal derivative
  coupling}}, \href{http://dx.doi.org/10.1103/PhysRevD.80.103505}{\emph{Phys.
  Rev. D} {\bfseries 80} (2009) 103505},
  [\href{https://arxiv.org/abs/0910.0980}{{\ttfamily 0910.0980}}].

\bibitem{Germani:2011mx}
C.~Germani, \emph{{Slow Roll Inflation: A Somehow Different Perspective}},
  {\emph{Rom. J. Phys.} {\bfseries 57} (2012) 841--848},
  [\href{https://arxiv.org/abs/1112.1083}{{\ttfamily 1112.1083}}].

\bibitem{Tsujikawa:2012mk}
S.~Tsujikawa, \emph{{Observational tests of inflation with a field derivative
  coupling to gravity}},
  \href{http://dx.doi.org/10.1103/PhysRevD.85.083518}{\emph{Phys. Rev. D}
  {\bfseries 85} (2012) 083518},
  [\href{https://arxiv.org/abs/1201.5926}{{\ttfamily 1201.5926}}].

\bibitem{Yang:2015pga}
N.~Yang, Q.~Fei, Q.~Gao and Y.~Gong, \emph{{Inflationary models with
  non-minimally derivative coupling}},
  \href{http://dx.doi.org/10.1088/0264-9381/33/20/205001}{\emph{Class. Quant.
  Grav.} {\bfseries 33} (2016) 205001},
  [\href{https://arxiv.org/abs/1504.05839}{{\ttfamily 1504.05839}}].

\bibitem{Fu:2019ttf}
C.~Fu, P.~Wu and H.~Yu, \emph{{Primordial Black Holes from Inflation with
  Nonminimal Derivative Coupling}},
  \href{http://dx.doi.org/10.1103/PhysRevD.100.063532}{\emph{Phys. Rev. D}
  {\bfseries 100} (2019) 063532},
  [\href{https://arxiv.org/abs/1907.05042}{{\ttfamily 1907.05042}}].

\bibitem{Sato:2020ghj}
S.~Sato and K.-i. Maeda, \emph{{Stability of hybrid Higgs inflation}},
  \href{http://dx.doi.org/10.1103/PhysRevD.101.103520}{\emph{Phys. Rev. D}
  {\bfseries 101} (2020) 103520},
  [\href{https://arxiv.org/abs/2001.00154}{{\ttfamily 2001.00154}}].

\bibitem{Germani:2010gm}
C.~Germani and A.~Kehagias, \emph{{New Model of Inflation with Non-minimal
  Derivative Coupling of Standard Model Higgs Boson to Gravity}},
  \href{http://dx.doi.org/10.1103/PhysRevLett.105.011302}{\emph{Phys. Rev.
  Lett.} {\bfseries 105} (2010) 011302},
  [\href{https://arxiv.org/abs/1003.2635}{{\ttfamily 1003.2635}}].

\bibitem{Germani:2010ux}
C.~Germani and A.~Kehagias, \emph{{Cosmological Perturbations in the New Higgs
  Inflation}},
  \href{http://dx.doi.org/10.1088/1475-7516/2010/05/019}{\emph{JCAP} {\bfseries
  05} (2010) 019}, [\href{https://arxiv.org/abs/1003.4285}{{\ttfamily
  1003.4285}}].

\bibitem{Germani:2014hqa}
C.~Germani, Y.~Watanabe and N.~Wintergerst, \emph{{Self-unitarization of New
  Higgs Inflation and compatibility with Planck and BICEP2 data}},
  \href{http://dx.doi.org/10.1088/1475-7516/2014/12/009}{\emph{JCAP} {\bfseries
  12} (2014) 009}, [\href{https://arxiv.org/abs/1403.5766}{{\ttfamily
  1403.5766}}].

\bibitem{DiVita:2015bha}
S.~Di~Vita and C.~Germani, \emph{{Electroweak vacuum stability and inflation
  via nonminimal derivative couplings to gravity}},
  \href{http://dx.doi.org/10.1103/PhysRevD.93.045005}{\emph{Phys. Rev. D}
  {\bfseries 93} (2016) 045005},
  [\href{https://arxiv.org/abs/1508.04777}{{\ttfamily 1508.04777}}].

\bibitem{Escriva:2016cwl}
A.~Escriv{\`a} and C.~Germani, \emph{{Beyond dimensional analysis: Higgs and
  new Higgs inflations do not violate unitarity}},
  \href{http://dx.doi.org/10.1103/PhysRevD.95.123526}{\emph{Phys. Rev.}
  {\bfseries D95} (2017) 123526},
  [\href{https://arxiv.org/abs/1612.06253}{{\ttfamily 1612.06253}}].

\bibitem{Fumagalli:2017cdo}
J.~Fumagalli, S.~Mooij and M.~Postma, \emph{{Unitarity and predictiveness in
  new Higgs inflation}},
  \href{http://dx.doi.org/10.1007/JHEP03(2018)038}{\emph{JHEP} {\bfseries 03}
  (2018) 038}, [\href{https://arxiv.org/abs/1711.08761}{{\ttfamily
  1711.08761}}].

\bibitem{Granda:2019wip}
L.~Granda, D.~Jimenez and W.~Cardona, \emph{{Higgs inflation with non-minimal
  derivative coupling to gravity}},
  \href{http://dx.doi.org/10.1016/j.astropartphys.2020.102459}{\emph{Astropart.
  Phys.} {\bfseries 121} (2020) 102459},
  [\href{https://arxiv.org/abs/1911.02901}{{\ttfamily 1911.02901}}].

\bibitem{Granda:2019wyi}
L.~N. Granda and D.~F. Jimenez, \emph{{Higgs Inflation with linear and
  quadratic curvature corrections}},
  \href{https://arxiv.org/abs/1910.11289}{{\ttfamily 1910.11289}}.

\bibitem{Fumagalli:2020ody}
J.~Fumagalli, M.~Postma and M.~Van Den~Bout, \emph{{Matching and running
  sensitivity in non-renormalizable inflationary models}},
  \href{http://dx.doi.org/10.1007/JHEP09(2020)114}{\emph{JHEP} {\bfseries 09}
  (2020) 114}, [\href{https://arxiv.org/abs/2005.05905}{{\ttfamily
  2005.05905}}].

\bibitem{Kobayashi:2010cm}
T.~Kobayashi, M.~Yamaguchi and J.~Yokoyama, \emph{{G-inflation: Inflation
  driven by the Galileon field}},
  \href{http://dx.doi.org/10.1103/PhysRevLett.105.231302}{\emph{Phys. Rev.
  Lett.} {\bfseries 105} (2010) 231302},
  [\href{https://arxiv.org/abs/1008.0603}{{\ttfamily 1008.0603}}].

\bibitem{Kamada:2010qe}
K.~Kamada, T.~Kobayashi, M.~Yamaguchi and J.~Yokoyama, \emph{{Higgs
  G-inflation}},
  \href{http://dx.doi.org/10.1103/PhysRevD.83.083515}{\emph{Phys. Rev. D}
  {\bfseries 83} (2011) 083515},
  [\href{https://arxiv.org/abs/1012.4238}{{\ttfamily 1012.4238}}].

\bibitem{Kobayashi:2011nu}
T.~Kobayashi, M.~Yamaguchi and J.~Yokoyama, \emph{{Generalized G-inflation:
  Inflation with the most general second-order field equations}},
  \href{http://dx.doi.org/10.1143/PTP.126.511}{\emph{Prog. Theor. Phys.}
  {\bfseries 126} (2011) 511--529},
  [\href{https://arxiv.org/abs/1105.5723}{{\ttfamily 1105.5723}}].

\bibitem{Kamada:2012se}
K.~Kamada, T.~Kobayashi, T.~Takahashi, M.~Yamaguchi and J.~Yokoyama,
  \emph{{Generalized Higgs inflation}},
  \href{http://dx.doi.org/10.1103/PhysRevD.86.023504}{\emph{Phys. Rev. D}
  {\bfseries 86} (2012) 023504},
  [\href{https://arxiv.org/abs/1203.4059}{{\ttfamily 1203.4059}}].

\bibitem{Kamada:2013bia}
K.~Kamada, T.~Kobayashi, T.~Kunimitsu, M.~Yamaguchi and J.~Yokoyama,
  \emph{{Graceful exit from Higgs $G$ inflation}},
  \href{http://dx.doi.org/10.1103/PhysRevD.88.123518}{\emph{Phys. Rev. D}
  {\bfseries 88} (2013) 123518},
  [\href{https://arxiv.org/abs/1309.7410}{{\ttfamily 1309.7410}}].

\bibitem{Kunimitsu:2015faa}
T.~Kunimitsu, T.~Suyama, Y.~Watanabe and J.~Yokoyama, \emph{{Large tensor mode,
  field range bound and consistency in generalized G-inflation}},
  \href{http://dx.doi.org/10.1088/1475-7516/2015/08/044}{\emph{JCAP} {\bfseries
  08} (2015) 044}, [\href{https://arxiv.org/abs/1504.06946}{{\ttfamily
  1504.06946}}].

\bibitem{Sato:2017qau}
S.~Sato and K.-i. Maeda, \emph{{Hybrid Higgs Inflation: The Use of Disformal
  Transformation}},
  \href{http://dx.doi.org/10.1103/PhysRevD.97.083512}{\emph{Phys. Rev. D}
  {\bfseries 97} (2018) 083512},
  [\href{https://arxiv.org/abs/1712.04237}{{\ttfamily 1712.04237}}].

\bibitem{Woodard:2006nt}
R.~P. Woodard, \emph{{Avoiding dark energy with 1/R modifications of gravity}},
  \href{http://dx.doi.org/10.1007/978-3-540-71013-4_14}{\emph{Lect. Notes
  Phys.} {\bfseries 720} (2007) 403--433},
  [\href{https://arxiv.org/abs/astro-ph/0601672}{{\ttfamily
  astro-ph/0601672}}].

\bibitem{Horndeski:1974wa}
G.~W. Horndeski, \emph{{Second-order scalar-tensor field equations in a
  four-dimensional space}},
  \href{http://dx.doi.org/10.1007/BF01807638}{\emph{Int. J. Theor. Phys.}
  {\bfseries 10} (1974) 363--384}.

\bibitem{Langlois:2018dxi}
D.~Langlois, \emph{{Dark energy and modified gravity in degenerate higher-order
  scalar\textendash{}tensor (DHOST) theories: A review}},
  \href{http://dx.doi.org/10.1142/S0218271819420069}{\emph{Int. J. Mod. Phys.
  D} {\bfseries 28} (2019) 1942006},
  [\href{https://arxiv.org/abs/1811.06271}{{\ttfamily 1811.06271}}].

\bibitem{Kobayashi:2019hrl}
T.~Kobayashi, \emph{{Horndeski theory and beyond: a review}},
  \href{http://dx.doi.org/10.1088/1361-6633/ab2429}{\emph{Rept. Prog. Phys.}
  {\bfseries 82} (2019) 086901},
  [\href{https://arxiv.org/abs/1901.07183}{{\ttfamily 1901.07183}}].

\bibitem{Langlois:2017mxy}
D.~Langlois, M.~Mancarella, K.~Noui and F.~Vernizzi, \emph{{Effective
  Description of Higher-Order Scalar-Tensor Theories}},
  \href{http://dx.doi.org/10.1088/1475-7516/2017/05/033}{\emph{JCAP} {\bfseries
  05} (2017) 033}, [\href{https://arxiv.org/abs/1703.03797}{{\ttfamily
  1703.03797}}].

\bibitem{Takahashi:2021ttd}
K.~Takahashi, H.~Motohashi and M.~Minamitsuji, \emph{{Invertible disformal
  transformations with higher derivatives}},
  \href{http://dx.doi.org/10.1103/PhysRevD.105.024015}{\emph{Phys. Rev. D}
  {\bfseries 105} (2022) 024015},
  [\href{https://arxiv.org/abs/2111.11634}{{\ttfamily 2111.11634}}].

\bibitem{Takahashi:2022mew}
K.~Takahashi, M.~Minamitsuji and H.~Motohashi, \emph{{Generalized disformal
  Horndeski theories: Cosmological perturbations and consistent matter
  coupling}}, \href{http://dx.doi.org/10.1093/ptep/ptac161}{\emph{PTEP}
  {\bfseries 2023} (2023) 013E01},
  [\href{https://arxiv.org/abs/2209.02176}{{\ttfamily 2209.02176}}].

\bibitem{Langlois:2015cwa}
D.~Langlois and K.~Noui, \emph{{Degenerate higher derivative theories beyond
  Horndeski: evading the Ostrogradski instability}},
  \href{http://dx.doi.org/10.1088/1475-7516/2016/02/034}{\emph{JCAP} {\bfseries
  02} (2016) 034}, [\href{https://arxiv.org/abs/1510.06930}{{\ttfamily
  1510.06930}}].

\bibitem{DeFelice:2018ewo}
A.~De~Felice, D.~Langlois, S.~Mukohyama, K.~Noui and A.~Wang,
  \emph{{Generalized instantaneous modes in higher-order scalar-tensor
  theories}}, \href{http://dx.doi.org/10.1103/PhysRevD.98.084024}{\emph{Phys.
  Rev. D} {\bfseries 98} (2018) 084024},
  [\href{https://arxiv.org/abs/1803.06241}{{\ttfamily 1803.06241}}].

\bibitem{DeFelice:2021hps}
A.~De~Felice, S.~Mukohyama and K.~Takahashi, \emph{{Nonlinear definition of the
  shadowy mode in higher-order scalar-tensor theories}},
  \href{http://dx.doi.org/10.1088/1475-7516/2021/12/020}{\emph{JCAP} {\bfseries
  12} (2021) 020}, [\href{https://arxiv.org/abs/2110.03194}{{\ttfamily
  2110.03194}}].

\bibitem{Joshi:2021azw}
P.~Joshi and S.~Panda, \emph{{Higher derivative scalar tensor theory in unitary
  gauge}}, \href{http://dx.doi.org/10.1088/1475-7516/2022/03/022}{\emph{JCAP}
  {\bfseries 03} (2022) 022},
  [\href{https://arxiv.org/abs/2111.11791}{{\ttfamily 2111.11791}}].

\bibitem{Joshi:2023otx}
P.~Joshi, S.~Panda and A.~Vidyarthi, \emph{{Ghost free theory in unitary gauge:
  a~new~candidate}},
  \href{http://dx.doi.org/10.1088/1475-7516/2023/07/051}{\emph{JCAP} {\bfseries
  07} (2023) 051}, [\href{https://arxiv.org/abs/2303.12464}{{\ttfamily
  2303.12464}}].

\bibitem{Takahashi:2023jro}
K.~Takahashi, M.~Minamitsuji and H.~Motohashi, \emph{{Effective description of
  generalized disformal theories}},
  \href{https://arxiv.org/abs/2304.08624}{{\ttfamily 2304.08624}}.

\bibitem{Bahamonde:2019shr}
S.~Bahamonde, K.~F. Dialektopoulos and J.~Levi~Said, \emph{{Can Horndeski
  Theory be recast using Teleparallel Gravity?}},
  \href{http://dx.doi.org/10.1103/PhysRevD.100.064018}{\emph{Phys. Rev. D}
  {\bfseries 100} (2019) 064018},
  [\href{https://arxiv.org/abs/1904.10791}{{\ttfamily 1904.10791}}].

\bibitem{Bahamonde:2022cmz}
S.~Bahamonde, G.~Trenkler, L.~G. Trombetta and M.~Yamaguchi, \emph{{Symmetric
  teleparallel Horndeski gravity}},
  \href{http://dx.doi.org/10.1103/PhysRevD.107.104024}{\emph{Phys. Rev. D}
  {\bfseries 107} (2023) 104024},
  [\href{https://arxiv.org/abs/2212.08005}{{\ttfamily 2212.08005}}].

\bibitem{einstein1925}
A.~Einstein, \emph{{Einheitliche Feldtheorie von Gravitation und
  Elektrizit\"{a}t}}, {\emph{Sitzungber.Preuss.Akad.Wiss.} {\bfseries 22}
  (1925) 414--419}.

\bibitem{ferraris1982}
M.~Ferraris, M.~Francaviglia and C.~Reina, \emph{{Variational formulation of
  general relativity from 1915 to 1925 ``{Palatini}'s method'' discovered by
  {Einstein} in 1925}}, {\emph{Gen.Rel.Grav.} {\bfseries 14} (1982) 243--254}.

\bibitem{Helpin:2019kcq}
T.~Helpin and M.~S. Volkov, \emph{{Varying the Horndeski Lagrangian within the
  Palatini approach}},
  \href{http://dx.doi.org/10.1088/1475-7516/2020/01/044}{\emph{JCAP} {\bfseries
  01} (2020) 044}, [\href{https://arxiv.org/abs/1906.07607}{{\ttfamily
  1906.07607}}].

\bibitem{Helpin:2019vrv}
T.~Helpin and M.~S. Volkov, \emph{{A metric-affine version of the Horndeski
  theory}}, \href{http://dx.doi.org/10.1142/S0217751X20400102}{\emph{Int. J.
  Mod. Phys. A} {\bfseries 35} (2020) 2040010},
  [\href{https://arxiv.org/abs/1911.12768}{{\ttfamily 1911.12768}}].

\bibitem{Dong:2021jtd}
Y.-Q. Dong and Y.-X. Liu, \emph{{Polarization modes of gravitational waves in
  Palatini-Horndeski theory}},
  \href{http://dx.doi.org/10.1103/PhysRevD.105.064035}{\emph{Phys. Rev. D}
  {\bfseries 105} (2022) 064035},
  [\href{https://arxiv.org/abs/2111.07352}{{\ttfamily 2111.07352}}].

\bibitem{Dong:2022cvf}
Y.-Q. Dong, Y.-Q. Liu and Y.-X. Liu, \emph{{Constraining
  Palatini\textendash{}Horndeski theory with gravitational waves after
  GW170817}},
  \href{http://dx.doi.org/10.1140/epjc/s10052-023-11861-9}{\emph{Eur. Phys. J.
  C} {\bfseries 83} (2023) 702},
  [\href{https://arxiv.org/abs/2211.12056}{{\ttfamily 2211.12056}}].

\bibitem{Bauer:2008}
F.~Bauer and D.~A. Demir, \emph{{Inflation with Non-Minimal Coupling: Metric
  versus Palatini Formulations}},
  \href{http://dx.doi.org/10.1016/j.physletb.2008.06.014}{\emph{Phys. Lett.}
  {\bfseries B665} (2008) 222--226},
  [\href{https://arxiv.org/abs/0803.2664}{{\ttfamily 0803.2664}}].

\bibitem{Bauer:2010}
F.~Bauer and D.~A. Demir, \emph{{Higgs-Palatini Inflation and Unitarity}},
  \href{http://dx.doi.org/10.1016/j.physletb.2011.03.042}{\emph{Phys. Lett.}
  {\bfseries B698} (2011) 425--429},
  [\href{https://arxiv.org/abs/1012.2900}{{\ttfamily 1012.2900}}].

\bibitem{Rasanen:2017}
S.~R{\"a}s{\"a}nen and P.~Wahlman, \emph{{Higgs inflation with loop corrections
  in the Palatini formulation}},
  \href{http://dx.doi.org/10.1088/1475-7516/2017/11/047}{\emph{JCAP} {\bfseries
  1711} (2017) 047}, [\href{https://arxiv.org/abs/1709.07853}{{\ttfamily
  1709.07853}}].

\bibitem{Racioppi:2017spw}
A.~Racioppi, \emph{{Coleman-Weinberg linear inflation: metric vs. Palatini
  formulation}},
  \href{http://dx.doi.org/10.1088/1475-7516/2017/12/041}{\emph{JCAP} {\bfseries
  12} (2017) 041}, [\href{https://arxiv.org/abs/1710.04853}{{\ttfamily
  1710.04853}}].

\bibitem{Markkanen:2017tun}
T.~Markkanen, T.~Tenkanen, V.~Vaskonen and H.~Veerm\"ae, \emph{{Quantum
  corrections to quartic inflation with a non-minimal coupling: metric vs.
  Palatini}},
  \href{http://dx.doi.org/10.1088/1475-7516/2018/03/029}{\emph{JCAP} {\bfseries
  03} (2018) 029}, [\href{https://arxiv.org/abs/1712.04874}{{\ttfamily
  1712.04874}}].

\bibitem{Enckell:2018a}
V.-M. Enckell, K.~Enqvist, S.~R{\"a}s{\"a}nen and E.~Tomberg, \emph{{Higgs
  inflation at the hilltop}},
  \href{http://dx.doi.org/10.1088/1475-7516/2018/06/005}{\emph{JCAP} {\bfseries
  1806} (2018) 005}, [\href{https://arxiv.org/abs/1802.09299}{{\ttfamily
  1802.09299}}].

\bibitem{Rasanen:2018fom}
S.~R{\"a}s{\"a}nen and E.~Tomberg, \emph{{Planck scale black hole dark matter
  from Higgs inflation}},
  \href{http://dx.doi.org/10.1088/1475-7516/2019/01/038}{\emph{JCAP} {\bfseries
  01} (2019) 038}, [\href{https://arxiv.org/abs/1810.12608}{{\ttfamily
  1810.12608}}].

\bibitem{Rasanen:2018ihz}
S.~R{\"a}s{\"a}nen, \emph{{Higgs inflation in the Palatini formulation with
  kinetic terms for the metric}},
  \href{http://dx.doi.org/10.21105/astro.1811.09514}{\emph{Open J. Astrophys.}
  {\bfseries 2} (2019) 1}, [\href{https://arxiv.org/abs/1811.09514}{{\ttfamily
  1811.09514}}].

\bibitem{Rubio:2019}
J.~Rubio and E.~S. Tomberg, \emph{{Preheating in Palatini Higgs inflation}},
  \href{http://dx.doi.org/10.1088/1475-7516/2019/04/021}{\emph{JCAP} {\bfseries
  1904} (2019) 021}, [\href{https://arxiv.org/abs/1902.10148}{{\ttfamily
  1902.10148}}].

\bibitem{Jinno:2019und}
R.~Jinno, M.~Kubota, K.-y. Oda and S.~C. Park, \emph{{Higgs inflation in metric
  and Palatini formalisms: Required suppression of higher dimensional
  operators}},
  \href{http://dx.doi.org/10.1088/1475-7516/2020/03/063}{\emph{JCAP} {\bfseries
  03} (2020) 063}, [\href{https://arxiv.org/abs/1904.05699}{{\ttfamily
  1904.05699}}].

\bibitem{Tenkanen:2020dge}
T.~Tenkanen, \emph{{Tracing the high energy theory of gravity: an introduction
  to Palatini inflation}},
  \href{http://dx.doi.org/10.1007/s10714-020-02682-2}{\emph{Gen. Rel. Grav.}
  {\bfseries 52} (2020) 33},
  [\href{https://arxiv.org/abs/2001.10135}{{\ttfamily 2001.10135}}].

\bibitem{Shaposhnikov:2020fdv}
M.~Shaposhnikov, A.~Shkerin and S.~Zell, \emph{{Quantum Effects in Palatini
  Higgs Inflation}},
  \href{http://dx.doi.org/10.1088/1475-7516/2020/07/064}{\emph{JCAP} {\bfseries
  07} (2020) 064}, [\href{https://arxiv.org/abs/2002.07105}{{\ttfamily
  2002.07105}}].

\bibitem{McDonald:2020lpz}
J.~McDonald, \emph{{Does Palatini Higgs Inflation Conserve Unitarity?}},
  \href{http://dx.doi.org/10.1088/1475-7516/2021/04/069}{\emph{JCAP} {\bfseries
  04} (2021) 069}, [\href{https://arxiv.org/abs/2007.04111}{{\ttfamily
  2007.04111}}].

\bibitem{Shaposhnikov:2020frq}
M.~Shaposhnikov, A.~Shkerin, I.~Timiryasov and S.~Zell, \emph{{Einstein-Cartan
  gravity, matter, and scale-invariant generalization~}},
  \href{http://dx.doi.org/10.1007/JHEP08(2021)162}{\emph{JHEP} {\bfseries 10}
  (2020) 177}, [\href{https://arxiv.org/abs/2007.16158}{{\ttfamily
  2007.16158}}].

\bibitem{Enckell:2020lvn}
V.-M. Enckell, S.~Nurmi, S.~R\"as\"anen and E.~Tomberg, \emph{{Critical point
  Higgs inflation in the Palatini formulation}},
  \href{http://dx.doi.org/10.1007/JHEP04(2021)059}{\emph{JHEP} {\bfseries 04}
  (2021) 059}, [\href{https://arxiv.org/abs/2012.03660}{{\ttfamily
  2012.03660}}].

\bibitem{Antoniadis:2021axu}
I.~Antoniadis, A.~Guillen and K.~Tamvakis, \emph{{Ultraviolet behaviour of
  Higgs inflation models}},
  \href{http://dx.doi.org/10.1007/JHEP05(2022)074}{\emph{JHEP} {\bfseries 08}
  (2021) 018}, [\href{https://arxiv.org/abs/2106.09390}{{\ttfamily
  2106.09390}}].

\bibitem{Mikura:2021clt}
Y.~Mikura and Y.~Tada, \emph{{On UV-completion of Palatini-Higgs inflation}},
  \href{http://dx.doi.org/10.1088/1475-7516/2022/05/035}{\emph{JCAP} {\bfseries
  05} (2022) 035}, [\href{https://arxiv.org/abs/2110.03925}{{\ttfamily
  2110.03925}}].

\bibitem{Ito:2021ssc}
A.~Ito, W.~Khater and S.~R{\"a}s{\"a}nen, \emph{{Tree-level unitarity in Higgs
  inflation in the metric and the Palatini formulation}},
  \href{http://dx.doi.org/10.1007/JHEP06(2022)164}{\emph{JHEP} {\bfseries 06}
  (2022) 164}, [\href{https://arxiv.org/abs/2111.05621}{{\ttfamily
  2111.05621}}].

\bibitem{Karananas:2022byw}
G.~K. Karananas, M.~Shaposhnikov and S.~Zell, \emph{{Field redefinitions,
  perturbative unitarity and Higgs inflation}},
  \href{http://dx.doi.org/10.1007/JHEP06(2022)132}{\emph{JHEP} {\bfseries 06}
  (2022) 132}, [\href{https://arxiv.org/abs/2203.09534}{{\ttfamily
  2203.09534}}].

\bibitem{Dux:2022kuk}
F.~Dux, A.~Florio, J.~Klari\'c, A.~Shkerin and I.~Timiryasov, \emph{{Preheating
  in Palatini Higgs inflation on the lattice}},
  \href{http://dx.doi.org/10.1088/1475-7516/2022/09/015}{\emph{JCAP} {\bfseries
  09} (2022) 015}, [\href{https://arxiv.org/abs/2203.13286}{{\ttfamily
  2203.13286}}].

\bibitem{Gialamas:2023flv}
I.~D. Gialamas, A.~Karam, T.~D. Pappas and E.~Tomberg, \emph{{Implications of
  Palatini gravity for inflation and beyond}},
  \href{https://arxiv.org/abs/2303.14148}{{\ttfamily 2303.14148}}.

\bibitem{Piani:2023aof}
M.~Piani and J.~Rubio, \emph{{Preheating in Einstein-Cartan Higgs Inflation:
  Oscillon formation}},  \href{https://arxiv.org/abs/2304.13056}{{\ttfamily
  2304.13056}}.

\bibitem{Poisson:2023tja}
A.~Poisson, I.~Timiryasov and S.~Zell, \emph{{Critical Points in Palatini Higgs
  Inflation with Small Non-Minimal Coupling}},
  \href{https://arxiv.org/abs/2306.03893}{{\ttfamily 2306.03893}}.

\bibitem{Gumjudpai:2016ioy}
N.~Kaewkhao and B.~Gumjudpai, \emph{{Cosmology of non-minimal derivative
  coupling to gravity in Palatini formalism and its chaotic inflation}},
  \href{http://dx.doi.org/10.1016/j.dark.2018.02.004}{\emph{Phys. Dark Univ.}
  {\bfseries 20} (2018) 20--27},
  [\href{https://arxiv.org/abs/1608.04014}{{\ttfamily 1608.04014}}].

\bibitem{Galtsov:2018xuc}
D.~Gal'tsov and S.~Zhidkova, \emph{{Ghost-free Palatini derivative
  scalar\textendash{}tensor theory: Desingularization and the speed test}},
  \href{http://dx.doi.org/10.1016/j.physletb.2019.01.061}{\emph{Phys. Lett. B}
  {\bfseries 790} (2019) 453--457},
  [\href{https://arxiv.org/abs/1808.00492}{{\ttfamily 1808.00492}}].

\bibitem{Gialamas:2020vto}
I.~D. Gialamas, A.~Karam, A.~Lykkas and T.~D. Pappas, \emph{{Palatini-Higgs
  inflation with nonminimal derivative coupling}},
  \href{http://dx.doi.org/10.1103/PhysRevD.102.063522}{\emph{Phys. Rev. D}
  {\bfseries 102} (2020) 063522},
  [\href{https://arxiv.org/abs/2008.06371}{{\ttfamily 2008.06371}}].

\bibitem{Dioguardi:2023jwa}
C.~Dioguardi and A.~Racioppi, \emph{{Palatini $F(R,X)$: a new framework for
  inflationary attractors}},
  \href{https://arxiv.org/abs/2307.02963}{{\ttfamily 2307.02963}}.

\bibitem{Luo:2014eda}
X.~Luo, P.~Wu and H.~Yu, \emph{{Non-minimal derivatively coupled quintessence
  in the Palatini formalism}},
  \href{http://dx.doi.org/10.1007/s10509-014-1795-0}{\emph{Astrophys. Space
  Sci.} {\bfseries 350} (2014) 831--837}.

\bibitem{Aoki:2018lwx}
K.~Aoki and K.~Shimada, \emph{{Galileon and generalized Galileon with
  projective invariance in a metric-affine formalism}},
  \href{http://dx.doi.org/10.1103/PhysRevD.98.044038}{\emph{Phys. Rev. D}
  {\bfseries 98} (2018) 044038},
  [\href{https://arxiv.org/abs/1806.02589}{{\ttfamily 1806.02589}}].

\bibitem{Aoki:2019rvi}
K.~Aoki and K.~Shimada, \emph{{Scalar-metric-affine theories: Can we get
  ghost-free theories from symmetry?}},
  \href{http://dx.doi.org/10.1103/PhysRevD.100.044037}{\emph{Phys. Rev.}
  {\bfseries D100} (2019) 044037},
  [\href{https://arxiv.org/abs/1904.10175}{{\ttfamily 1904.10175}}].

\bibitem{Galtsov:2020jnu}
D.~V. Gal'tsov, \emph{{Conformal and kinetic couplings as two Jordan frames of
  the same theory: Conformal and kinetic couplings}},
  \href{http://dx.doi.org/10.1140/epjc/s10052-020-8017-4}{\emph{Eur. Phys. J.
  C} {\bfseries 80} (2020) 443},
  [\href{https://arxiv.org/abs/2001.03221}{{\ttfamily 2001.03221}}].

\bibitem{BeltranJimenez:2020sqf}
J.~Beltr\'an~Jim\'enez and A.~Delhom, \emph{{Instabilities in metric-affine
  theories of gravity with higher order curvature terms}},
  \href{http://dx.doi.org/10.1140/epjc/s10052-020-8143-z}{\emph{Eur. Phys. J.
  C} {\bfseries 80} (2020) 585},
  [\href{https://arxiv.org/abs/2004.11357}{{\ttfamily 2004.11357}}].

\bibitem{BeltranJimenez:2019acz}
J.~Beltrán~Jiménez and A.~Delhom, \emph{{Ghosts in metric-affine higher order
  curvature gravity}},
  \href{http://dx.doi.org/10.1140/epjc/s10052-019-7149-x}{\emph{Eur. Phys. J.
  C} {\bfseries 79} (2019) 656},
  [\href{https://arxiv.org/abs/1901.08988}{{\ttfamily 1901.08988}}].

\bibitem{Annala:2022gtl}
J.~Annala and S.~R{\"a}s{\"a}nen, \emph{{Stability of non-degenerate Ricci-type
  Palatini theories}},
  \href{http://dx.doi.org/10.1088/1475-7516/2023/04/014}{\emph{JCAP} {\bfseries
  04} (2023) 014}, [\href{https://arxiv.org/abs/2212.09820}{{\ttfamily
  2212.09820}}].

\bibitem{Minamitsuji:2014waa}
M.~Minamitsuji, \emph{{Disformal transformation of cosmological
  perturbations}},
  \href{http://dx.doi.org/10.1016/j.physletb.2014.08.037}{\emph{Phys. Lett. B}
  {\bfseries 737} (2014) 139--150},
  [\href{https://arxiv.org/abs/1409.1566}{{\ttfamily 1409.1566}}].

\bibitem{Tsujikawa:2014uza}
S.~Tsujikawa, \emph{{Disformal invariance of cosmological perturbations in a
  generalized class of Horndeski theories}},
  \href{http://dx.doi.org/10.1088/1475-7516/2015/04/043}{\emph{JCAP} {\bfseries
  04} (2015) 043}, [\href{https://arxiv.org/abs/1412.6210}{{\ttfamily
  1412.6210}}].

\bibitem{Watanabe:2015uqa}
Y.~Watanabe, A.~Naruko and M.~Sasaki, \emph{{Multi-disformal invariance of
  non-linear primordial perturbations}},
  \href{http://dx.doi.org/10.1209/0295-5075/111/39002}{\emph{EPL} {\bfseries
  111} (2015) 39002}, [\href{https://arxiv.org/abs/1504.00672}{{\ttfamily
  1504.00672}}].

\bibitem{Motohashi:2015pra}
H.~Motohashi and J.~White, \emph{{Disformal invariance of curvature
  perturbation}},
  \href{http://dx.doi.org/10.1088/1475-7516/2016/02/065}{\emph{JCAP} {\bfseries
  02} (2016) 065}, [\href{https://arxiv.org/abs/1504.00846}{{\ttfamily
  1504.00846}}].

\bibitem{Domenech:2015hka}
G.~Domènech, A.~Naruko and M.~Sasaki, \emph{{Cosmological disformal
  invariance}},
  \href{http://dx.doi.org/10.1088/1475-7516/2015/10/067}{\emph{JCAP} {\bfseries
  10} (2015) 067}, [\href{https://arxiv.org/abs/1505.00174}{{\ttfamily
  1505.00174}}].

\bibitem{Chiba:2020mte}
T.~Chiba, F.~Chibana and M.~Yamaguchi, \emph{{Disformal invariance of
  cosmological observables}},
  \href{http://dx.doi.org/10.1088/1475-7516/2020/06/003}{\emph{JCAP} {\bfseries
  06} (2020) 003}, [\href{https://arxiv.org/abs/2003.10633}{{\ttfamily
  2003.10633}}].

\bibitem{Bettoni:2013diz}
D.~Bettoni and S.~Liberati, \emph{{Disformal invariance of second order
  scalar-tensor theories: Framing the Horndeski action}},
  \href{http://dx.doi.org/10.1103/PhysRevD.88.084020}{\emph{Phys. Rev. D}
  {\bfseries 88} (2013) 084020},
  [\href{https://arxiv.org/abs/1306.6724}{{\ttfamily 1306.6724}}].

\bibitem{Zumalacarregui:2013pma}
M.~Zumalacárregui and J.~García-Bellido, \emph{{Transforming gravity: from
  derivative couplings to matter to second-order scalar-tensor theories beyond
  the Horndeski Lagrangian}},
  \href{http://dx.doi.org/10.1103/PhysRevD.89.064046}{\emph{Phys. Rev. D}
  {\bfseries 89} (2014) 064046},
  [\href{https://arxiv.org/abs/1308.4685}{{\ttfamily 1308.4685}}].

\bibitem{Fumagalli:2016afy}
J.~Fumagalli, S.~Mooij and M.~Postma, \emph{{Disformal transformations as a
  change of units}},  \href{https://arxiv.org/abs/1610.08460}{{\ttfamily
  1610.08460}}.

\bibitem{Takahashi:2017zgr}
K.~Takahashi, H.~Motohashi, T.~Suyama and T.~Kobayashi, \emph{{General
  invertible transformation and physical degrees of freedom}},
  \href{http://dx.doi.org/10.1103/PhysRevD.95.084053}{\emph{Phys. Rev. D}
  {\bfseries 95} (2017) 084053},
  [\href{https://arxiv.org/abs/1702.01849}{{\ttfamily 1702.01849}}].

\bibitem{Magnano:1987zz}
G.~Magnano, M.~Ferraris and M.~Francaviglia, \emph{{Nonlinear gravitational
  Lagrangians}}, \href{http://dx.doi.org/10.1007/BF00760651}{\emph{Gen. Rel.
  Grav.} {\bfseries 19} (1987) 465}.

\bibitem{Koga:1998un}
J.-i. Koga and K.-i. Maeda, \emph{{Equivalence of black hole thermodynamics
  between a generalized theory of gravity and the Einstein theory}},
  \href{http://dx.doi.org/10.1103/PhysRevD.58.064020}{\emph{Phys. Rev. D}
  {\bfseries 58} (1998) 064020},
  [\href{https://arxiv.org/abs/gr-qc/9803086}{{\ttfamily gr-qc/9803086}}].

\bibitem{Sotiriou:2008}
T.~P. Sotiriou and V.~Faraoni, \emph{{f(R) Theories Of Gravity}},
  \href{http://dx.doi.org/10.1103/RevModPhys.82.451}{\emph{Rev. Mod. Phys.}
  {\bfseries 82} (2010) 451--497},
  [\href{https://arxiv.org/abs/0805.1726}{{\ttfamily 0805.1726}}].

\bibitem{Afonso:2017}
V.~I. Afonso, C.~Bejarano, J.~{Beltran Jimenez}, G.~J. Olmo and E.~Orazi,
  \emph{{The trivial role of torsion in projective invariant theories of
  gravity with non-minimally coupled matter fields}},
  \href{http://dx.doi.org/10.1088/1361-6382/aa9151}{\emph{Class. Quant. Grav.}
  {\bfseries 34} (2017) 235003},
  [\href{https://arxiv.org/abs/1705.03806}{{\ttfamily 1705.03806}}].

\bibitem{xAct}
J.~M. Mart\'in-Garc\'ia, ``{xAct, {Efficient} tensor computer algebra for the
  {Wolfram} {Language}}.'' \url{http://www.xact.es/}.

\end{thebibliography}\endgroup


\end{document}